\documentclass[12pt,a4paper]{iopart}

\usepackage{iopams}
\usepackage{setstack}
\usepackage{graphicx}
\usepackage{multicol}
\usepackage{color}
\usepackage{hyperref}

\usepackage{tikz}
\usetikzlibrary{decorations.markings}
\usetikzlibrary{decorations.pathreplacing}

\newcommand{\sgn}{\mathrm{sgn}}

\renewcommand{\Re}{\mathrm{Re}}
\renewcommand{\Im}{\mathrm{Im}}
\renewcommand{\[}{[\![}
\renewcommand{\]}{]\!]}

\newcommand{\keywords}[1]{\noindent\textbf{Keywords:} #1}

\renewcommand{\text}[1]{\mathrm{#1}}

\begin{document}

\title{Spectral gaps of open TASEP in the maximal current phase}
\author{Ulysse Godreau, Sylvain Prolhac}
\address{Laboratoire de Physique Th\'eorique, IRSAMC, UPS, Universit\'e de Toulouse, France}
\date{}


\begin{abstract}
We study spectral gaps of the one-dimensional totally asymmetric simple exclusion process (TASEP) with open boundaries in the maximal current phase. Earlier results for the model with periodic boundaries suggest that the gaps contributing to the universal KPZ regime may be understood as points on an infinite genus Riemann surface built from a parametric representation of the cumulant generating function of the current. We perform explicit analytic continuations from the known large deviations of the current for open TASEP, and confirm the results for the gaps by an exact Bethe ansatz calculation, with additional checks using high precision extrapolation numerics.

\vspace{5mm}

\end{abstract}

\begin{section}{Introduction}
The one-dimensional totally asymmetric simple exclusion process (TASEP) \cite{D1998.1,S2001.1,PS2002.1,GM2006.1} is a Markov process featuring hard-core particles hopping locally from any site $i$ to the next site $i+1$ of a one-dimensional lattice. TASEP is the archetype of a N-body stochastic process breaking detailed balance, with non-trivial interactions leading to a rich large scale behaviour, and may be used to model some aspects of protein synthesis \cite{CL2004.1,ZJS2011.1}, transport in cells \cite{CSN2005.1,CMZ2011.1,BN2013.1} or even road traffic \cite{CSS2000.1,H2001.1,AFS2017.1}.

We consider in this paper TASEP on a finite interval of $L$ sites, with a hopping rate equal to $1$ in the bulk of the system (i.e. a particle at site $i$ hops to the next site $i+1$ with probability $\rmd t$ in a small time interval $\rmd t$), injection of particles at site $1$ with rate $\alpha$ and removal of particles from site $L$ with rate $\beta$. This model exhibits boundary induced phase transitions, with three phases \cite{DEHP1993.1} (called low density, high density, and called maximal current phase) for the average density of particles in the stationary state depending on the values of $\alpha$ and $\beta$.

At large scales, the discrete occupation numbers of sites on the lattice can be replaced by a continuous density field. The deterministic evolution of this density field is given by Burgers' equation \cite{S1991.1,F2018.1} on the hydrodynamic time scale $t\sim L$. Fluctuations of the density field depend on the region considered in the phase diagram. In the maximal current phase $\alpha,\beta>1/2$, which is the focus of this paper, fluctuations belong to KPZ universality \cite{KPZ1986.1,HHZ1995.1,S2006.1,KK2010.1,HHT2015.1,S2019.1}, with a correlation length growing as $t^{2/3}$ before saturating at $L$ when $t\gg L^{3/2}$. The crossover scale $t\sim L^{3/2}$ corresponds to the KPZ fixed point in finite volume with appropriate boundary conditions \cite{CS2018.1,P2019.1,GPS2020.1}, and describes the relaxation process from KPZ fluctuations on $\mathbb{R}$ \cite{C2011.1} at short time to a Brownian stationary state \cite{DEL2004.1,BW2019.1} with non-Gaussian large deviations \cite{GLMV2012.1}.

The characteristic time for the relaxation of KPZ fluctuations to the stationary state is governed by the spectral gap $E_{1}$ of TASEP, which scales as $E_{1}\simeq e_{1}/L^{3/2}$ in the maximal current phase. For the model with periodic boundary conditions, the prefactor $e_{1}$, which was computed exactly by Gwa and Spohn \cite{GS1992.1,GS1992.2} almost 30 years ago, see also \cite{GM2005.1}, is expressed in terms of polylogarithms with half-integer index. For the model with open boundaries, the spectral gap was obtained by de Gier and Essler \cite{dGE2005.1,dGE2006.1} 15 years ago in the low and high density phases, and on the coexistence line between, see also \cite{dGE2008.1,dGFS2011.1} for the open exclusion process with partial asymmetry. An exact expression was missing so far for the spectral gap of open TASEP in the maximal current phase, which is relevant for KPZ universality. In this paper, we compute exactly $e_{1}$ for the maximal current phase, as $e_{1}=\chi_{\{-1,1\}}(v)/4$ where $v$ is solution of $\chi_{\{-1,1\}}'(v)=0$ and with the function $\chi_{\{-1,1\}}$ defined in (\ref{gap chi}). This is our main result. Additionally, we also compute the infinitely many higher gaps $e_{n}$ characterizing the relaxation process on the KPZ scale $t\sim L^{3/2}$.

The spectral gaps of TASEP in the maximal current phase are obtained in our paper in two different ways. The first one was suggested by the observation \cite{P2020.1} that the first gaps of TASEP with periodic boundary conditions are related to one another by analytic continuation with respect to (a parameter conjugate to) a fugacity counting the current of particles. Starting with the expression obtained by Lazarescu and Mallick \cite{LM2011.1,GLMV2012.1,L2015.1} for the stationary large deviation of the current, analytic continuation across branch cuts lead us to putative exact expressions for the spectral gaps, which were checked numerically with good agreement against the actual gaps of TASEP. This first approach confirms that the spectral gaps of open TASEP in the maximal current phase can be described by a Riemann surface $\mathcal{R}$ analogous to the one studied in \cite{P2020.1} for the model with periodic boundaries.

Our second approach to the spectral gaps in the maximal current phase is an actual Bethe ansatz calculation. Unlike TASEP with periodic boundaries, for which Bethe ansatz works in essentially the same way as for the Heisenberg quantum spin chain studied by Bethe 90 years ago \cite{B1931.1}, Bethe equations for integrable models with open boundaries have long been missing due to the absence of a reference state in the algebraic Bethe ansatz formalism. Breakthroughs papers by Cao et al. \cite{CLSW2003.1,CYSW2013.1,WYCS2015.1}, see also \cite{BC2013.1}, first for special values of boundary parameters, then for general values, have allowed to obtain Bethe equations \cite{dGE2005.1,dGE2006.1,LP2014.1,WYCCY2015.1} and eigenvectors \cite{C2015.1} for TASEP with open boundaries, see also \cite{CRS2011.1,CRV2014.1} for related works on the integrability of open exclusion processes. Different equivalent versions of the Bethe equations exist: very recently, Cramp\'e and Nepomechie \cite{CN2018.1} found new Bethe equations for open TASEP which have the same decoupling property as with periodic boundaries, and which are the starting point for our calculation of the gaps using Bethe ansatz. Our result matches with the one by analytic continuation, which confirms that excitations at the KPZ fixed point with open boundaries can be understood in a unified way as points on the Riemann surface $\mathcal{R}$ mentioned above.

In section~\ref{section main results}, we present our main results for the spectral gaps of open TASEP in the maximal current phase contributing to the relaxation of KPZ fluctuations. In section~\ref{section a.c.}, we study analytic continuations of the large deviation function of the current from \cite{GLMV2012.1} in the KPZ scaling, suggesting exact expressions for the spectral gaps. These expressions are then confirmed by a direct derivation from Bethe ansatz in section~\ref{section Bethe ansatz}, using the recent formulation obtained in \cite{CN2018.1}.
\end{section}

\begin{section}{Main results}
\label{section main results}
In this section, after recalling known facts about open TASEP, we summarize our main result for the spectral gaps corresponding to relaxation times of current fluctuations at the KPZ fixed point with open boundaries.

\begin{subsection}{Open TASEP}
TASEP in one dimension is a Markov process defined on a discrete chain on which hard-core particles hop between nearest neighbours in the same direction. In the open case which we consider, the chain of $L$ sites is in contact at both ends with particle resevoirs. The system evolves in time according to the following rules, see figure~\ref{dynamic_rules_tasep}: a particle on any site $i$ may hop to the site $i+1$ with rate $1$ (i.e. with probability $\rmd t$ during a small time interval $\rmd t$) if the latter site is empty. A particle may enter the system at site $1$, if empty, with rate $\alpha$. A particle at site $L$ may leave the system with rate $\beta$.

Depending on the values of the rates $\alpha$ and $\beta$, the system finds itself in either of three phases characterized in particular by different values of the stationary current and density of particle in the chain, see figure \ref{phase_diagram}. In the low and high density phases, the dynamics is essentially controlled by boundary effects. On the other hand, the maximal current phase, which is the focus of this paper, corresponds to an effective dynamics where all sites are coupled, and the corresponding height function is described by the KPZ fixed point on the interval, with an infinite slope at both ends \cite{CS2018.1,P2019.1,GPS2020.1}.

\begin{figure}
\begin{center}
\includegraphics{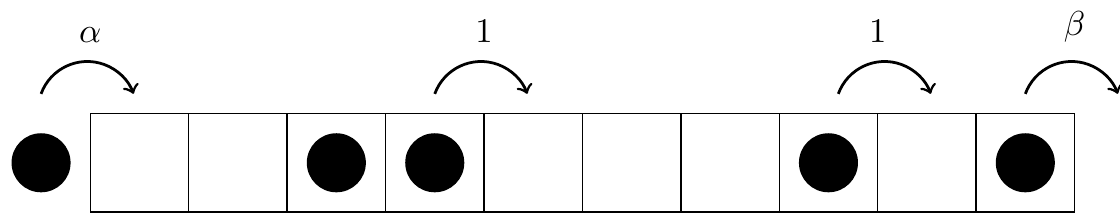}
\end{center}
\caption{Dynamical rules for TASEP with open boundaries.}
\label{dynamic_rules_tasep}
\end{figure}

The probability $P_t(\mathcal{C})$ of finding the system at time $t$ in the configuration $\mathcal{C}$ satisfies the master equation
\begin{equation}
\frac{d}{dt} P_t (\mathcal{C}) = \sum_{\mathcal{C}'} M_{\mathcal{C}, \mathcal{C}'} P_t (\mathcal{C}') \;,
\end{equation}
where $M$ is the Markov matrix of the process. If $\mathcal{C} \neq \mathcal{C}'$, $M_{\mathcal{C},\mathcal{C}'}$ is the rate of the transition $\mathcal{C}' \rightarrow \mathcal{C}$, and $M_{\mathcal{C},\mathcal{C}}= - \sum_{\mathcal{C}' \neq \mathcal{C}} P_t(\mathcal{C}')$ by conservation of the total probability. Each site can be described by a 2-dimensional space $\mathbb{C}^2$ with base vectors $|0\rangle,|1\rangle$ representing respectively empty and occupied state. The full Markov matrix acts on $(\mathbb{C}^2)^{\otimes L}$ and can be written as
\begin{equation}
 M = B_1 + \sum_{k=1}^{L-1} W_{k,k+1} + D_L \;,
\end{equation}
where the subscripts indicate on which sites the local update operators act non-trivially, with local matrices in canonical basis
\begin{equation}
\fl\hspace{15mm}
B_1=\left( \begin{array}{cc}
 -\alpha & 0 \\
 \alpha & 0
\end{array} \right)
\qquad
W=\left( \begin{array}{cccc}
 0 & 0 & 0 & 0 \\
 0 & 0 & 1 & 0 \\
 0 & 0 & -1 & 0 \\
 0 & 0 & 0 & 0
\end{array} \right)
\qquad
D_L=\left( \begin{array}{cc}
0 & \beta\\
0 & -\beta
\end{array} \right)\;.
\end{equation}

Defining the total number $Q_t$ of particle having entered the system since $t=0$, and the number $N_t$ of particles in the system at time $t$,the average stationary current $J$ and density of particles $\rho$ are given by
\begin{eqnarray}
J=\lim_{t \rightarrow \infty} \frac{\langle Q_t \rangle}{t}\\
\rho=\lim_{t \rightarrow \infty} \frac{\langle N_t \rangle}{L} \;.
\end{eqnarray}
Both $J$ and $\rho$ depend in principle on the system size $L$ and on the boundary rates $\alpha$ and $\beta$. At large $L$, they converge to simple values that depend on $\alpha,\beta$ only through the phase considered, see figure~\ref{phase_diagram}.

\begin{figure}
\begin{center}
\includegraphics{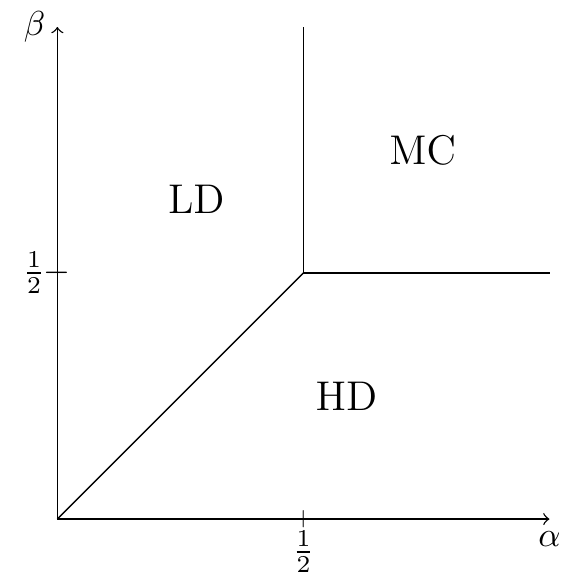}
\end{center}
\caption{Phase diagram of open TASEP. The low density phase (LD; $\rho=1-\beta$, $J=\beta(1-\beta)$) is separated from the high density phase (HD; $\rho=\alpha$, $J=\alpha(1-\alpha)$) by a first order transition at the coexistence line $\alpha=\beta<1/2$. The maximal current phase (MC; $\rho=\frac{1}{2}$, $J=\frac{1}{4}$) is separated from the LD and HD phases by second order transitions at $\alpha=1/2<\beta$ and $\beta=1/2<\alpha$.}
\label{phase_diagram}
\end{figure}

Probabilities $P_t(\mathcal{C})$ at a given time $t$ are not enough to compute the statistics of the current, and one has to consider instead joint probabilities $P_t(\mathcal{C}, Q)$ of the configuration and of the current. Introducing a fugacity $\gamma$ for the current, the quantity $F_t(\mathcal{C})= \sum_{Q=0}^{\infty}\rme^{\gamma Q} P_t(\mathcal{C}, Q)$ verifies the deformed master equation \cite{DL1998.1}
\begin{equation}
\frac{d}{dt} F_t (\mathcal{C}) = \sum_{\mathcal{C}'} M(\gamma)_{\mathcal{C}, \mathcal{C}'} F_t (\mathcal{C}') \;,
\end{equation}
where $M(\gamma)$ is obtained by multiplying by $\rme^{\gamma}$ the non-diagonal elements of $M$ corresponding to a transition where a particle enters the system. The stationary eigenvalue $E_0(\gamma)$ of $M(\gamma)$, which has the largest real part, is the cumulant generating function of the current in the stationary state:
\begin{equation}
E_0(\gamma) = \lim_{t \to \infty} \frac{1}{t} \log \langle \rme^{\gamma Q_t} \rangle \;.
\end{equation}
Higher eigenvalues $E_n(\gamma)$ give the relaxation times of eigenmodes for the generating function of the current at finite times:
\begin{equation}
\langle \rme^{\gamma Q_t} \rangle = \sum_{n}\theta_n \, \rme^{t E_n(\gamma)}\;,
\end{equation}
where the sum runs over all $2^L$ eigenstates, and with coefficients $\theta_n$ expressed as overlaps of the corresponding eigenstate with initial and final states.

In the following, we obtain exact expressions for the eigenvalues $E_n(\gamma)$ such that $E_n(\gamma)-E_0(\gamma)$ scales as $L^{-3/2}$ for large $L$ with fixed $\gamma\sqrt{L}$ in the maximal current phase, which are the ones contributing to the KPZ regime in finite volume. We show in particular that the gaps $E_n(\gamma)$ may be obtained from $E_0(\gamma)$ by analytic continuations.
\end{subsection}

\begin{subsection}{Exact expression of the first eigenvalues in the maximal current phase}
In the maximal current phase $\alpha, \beta>1/2$, the stationary eigenvalue $E_{0}(\gamma)$ of the deformed Markov matrix $M(\gamma)$ has at large $L$ the parametric expression \cite{GLMV2012.1}
\begin{eqnarray}
&& \gamma \simeq \frac{2\,\chi'(v)}{\sqrt{L}}\;,\\
&& E_{0} - \Big( \frac{1}{4} + \frac{1}{4L} \Big) \gamma \simeq \frac{\chi(v)}{4\,L^{3/2}}\;,
\end{eqnarray}
where the function $\chi$ is defined for $\Re~v<0$ by
\begin{equation}
\chi(v)=-\frac{1}{4\sqrt{\pi}}\sum_{k=1}^{\infty}\frac{(2k)!}{k!\,k^{k+5/2}}(-\rme^{v+1}/4)^{k}\;.
\end{equation}
Alternatively, the function $\chi$ has the integral expression
\begin{equation}
\label{chi_integral}
\chi(v)= \frac{1}{3\pi} \int_{-\infty}^{\infty} \rmd y \, \frac{(1-y^2)(3-y^2)}{1+y^{-2}\,\rme^{y^2-v-1}}
\end{equation}
when $\Re~v<0$ or $\Re~v\geq0$ and $-\pi<\Im~v<\pi$.

We claim that all the other eigenvalues in the maximal current phase $\alpha, \beta>1/2$ contributing to the KPZ regime have the similar parametric expression at large $L$:
\begin{eqnarray}
\label{vp_eq1}
&& \gamma \simeq \frac{2\,\chi_P'(v)}{\sqrt{L}}\;,\\
\label{vp_eq2}
&& E_P - \Big( \frac{1}{4} + \frac{1}{4L} \Big) \gamma \simeq \frac{\chi_P(v)}{4\,L^{3/2}}\;,
\end{eqnarray}
where $P$ is a finite set of integers labelling the eigenstates, see below, and $\chi_P$ is a branch of the multivalued function obtained by analytic continuations of $\chi$ above. Explicit expressions for the branches $\chi_P$, obtained in section~\ref{section a.c.} by analytic continuations and confirmed in section~\ref{bethe_ansatz_derivation} by a direct Bethe ansatz derivation, are given below, first for the gap, then for higher excited states.
\end{subsection}

\begin{subsection}{Spectral gap in the maximal current phase}
The spectral gap, corresponding to the eigenvalue of $M(\gamma)$ with largest real part after the stationary eigenvalue, has the parametric expression (\ref{vp_eq1}), (\ref{vp_eq2}) with label $P=\{-1,1\}$, and where $\chi_{\{-1,1\}}$ and its derivative are defined in terms of $\chi$ from (\ref{chi_integral}) by
\begin{eqnarray}
\label{gap chi}
&& \chi_{\{-1,1\}}(v)=\chi(v)+2\,\Big(\frac{(W_{1}(\rme^{-1-v}))^{3/2}}{3}+\sqrt{W_{1}(\rme^{-1-v})}\Big)\\
&&\hspace{31mm} +2\,\Big(\frac{(W_{-1}(\rme^{-1-v}))^{3/2}}{3}+\sqrt{W_{-1}(\rme^{-1-v})}\Big)\nonumber\\
\label{gap chi'}
&& \chi_{\{-1,1\}}'(v)=\chi'(v)-\sqrt{W_{1}(\rme^{-1-v})}-\sqrt{W_{-1}(\rme^{-1-v})}\;.
\end{eqnarray}
Here, $W_j$, $j\in\mathbb{Z}$ is the branch of the Lambert function $W_j(z)\,\rme^{W_j(z)}=z$ defined as the unique solution of $W_j(z)+\log W_j(z)=\log z + 2 \rmi \pi j$, with $\log$ the complex logarithm with branch cut $\mathbb{R}^-$, and the square root and power $3/2$ are defined as usual with branch cut $\mathbb{R}^-$.

\begin{table}
\begin{center}
$\begin{array}{|r|l|l|}
\hline
L & \quad e_1(L) & \quad e_1~~\text{extrapolated}\\\hline
3 & -3.15590 & -0.\\
4 & -3.31254 & -0.\\
5 & -3.39823 & -4.\\
6 & -3.44972 & -4.\\
7 & -3.48286 & -3.58\\
8 & -3.50533 & -3.578\\
9 & -3.52119 & -3.5780\\
10 & -3.53277 & -3.5781\\
&\quad\cdots&\qquad\cdots\\
15 & -3.56073 & -3.578064664\\
&\quad\cdots&\qquad\cdots\\
20 & -3.57038 & -3.5780646644761799\\
&\quad\cdots&\qquad\cdots\\
25 & -3.57460 & -3.5780646644761798418219\\
&\quad\cdots&\qquad\cdots\\
30 & -3.57672 & -3.57806466447617984182187036\\
&\quad\cdots&\qquad\cdots\\
35 & -3.57787 & -3.57806466447617984182187035823267\\
&\quad\cdots&\qquad\cdots\\
40 & -3.57854 & -3.5780646644761798418218703582326693543\\\hline
\end{array}$
\end{center}
\caption{Numerical evaluations of the spectral gap $e_{1}(L)=L^{3/2}E_{1}$ for $\alpha=\beta=1$ and a fugacity $\gamma=0$. The middle column corresponds to a direct numerical solution of the Bethe equations for a system of $L$ sites. The last column is the extrapolated value for all system sizes between $L=3$ and the value of $L$ for the current line using the BST algorithm with exponent $\theta=1$, see \ref{apendix num}. Numerical solutions of the Bethe equations are performed with 100 significant digits. Extrapolated values are truncated at the estimated order of magnitude of the error given by the BST algorithm.}
\label{gap_extrapolation}
\end{table}

For a fugacity $\gamma=0$, solving numerically (\ref{vp_eq1}) - (\ref{vp_eq2}), the spectral gap is equal in the thermodynamic limit to $E_1 \simeq e_1 L^{-3/2}$ with $e_1\approx-3.578064664476179841821870$, which matches with the value $e_1\approx-3.578$ obtained earlier by de Gier and Essler in \cite{dGE2006.1} from Bethe ansatz numerics. Additionally, our exact result agrees perfectly with high precision extrapolation of Bethe ansatz numerics, see table~\ref{gap_extrapolation}. The corresponding parameter $v$ solution of $\chi_{\{-1,1\}}'(v)=0$ is $v\approx4.228234$.
\end{subsection}

\begin{subsection}{Higher gaps in the maximal current phase}
Higher gaps of TASEP are again given by (\ref{vp_eq1})-(\ref{vp_eq2}) with
\begin{equation}
\label{chi_P}
\chi_P(v)=\chi_\emptyset (v) + 2 \rmi \sum_{j \in P} \eta_j(v) \;,
\end{equation}
where $P$ is a finite set of integers and the functions $\chi_\emptyset$ and $\eta_j$ are defined in section~\ref{section a.c.}. Numerical values for the first gaps with $\gamma=0$ are given in table~\ref{table_eigenstates}.

For values of $v$ with $-\pi<\Im~v<\pi$ one has in particular $\chi_\emptyset(v)=\chi(v)$ defined in (\ref{chi_integral}) and
\begin{eqnarray}
&& \rmi\eta_j(v) =\frac{(W_{j}(\rme^{-1-v}))^{3/2}}{3}+\sqrt{W_{j}(\rme^{-1-v})}\\
&& \rmi\eta_j'(v) =-\frac{1}{2}\,\sqrt{W_{j}(\rme^{-1-v})}
\end{eqnarray}
with the same notations as in (\ref{gap chi}), (\ref{gap chi'}), which is sufficient for the first few excited states when $\gamma\geq0$. For higher excited states, even when $\gamma\geq0$, the parameter $v$ may not lie in the range $-\pi<\Im~v<\pi$, and one has to use instead the general expressions (\ref{chi0}) for $\chi_\emptyset (v)$ and (\ref{etaj}), (\ref{yj}) for $\eta_j(v)$.

\begin{figure}
\begin{center}
\includegraphics[width=75mm]{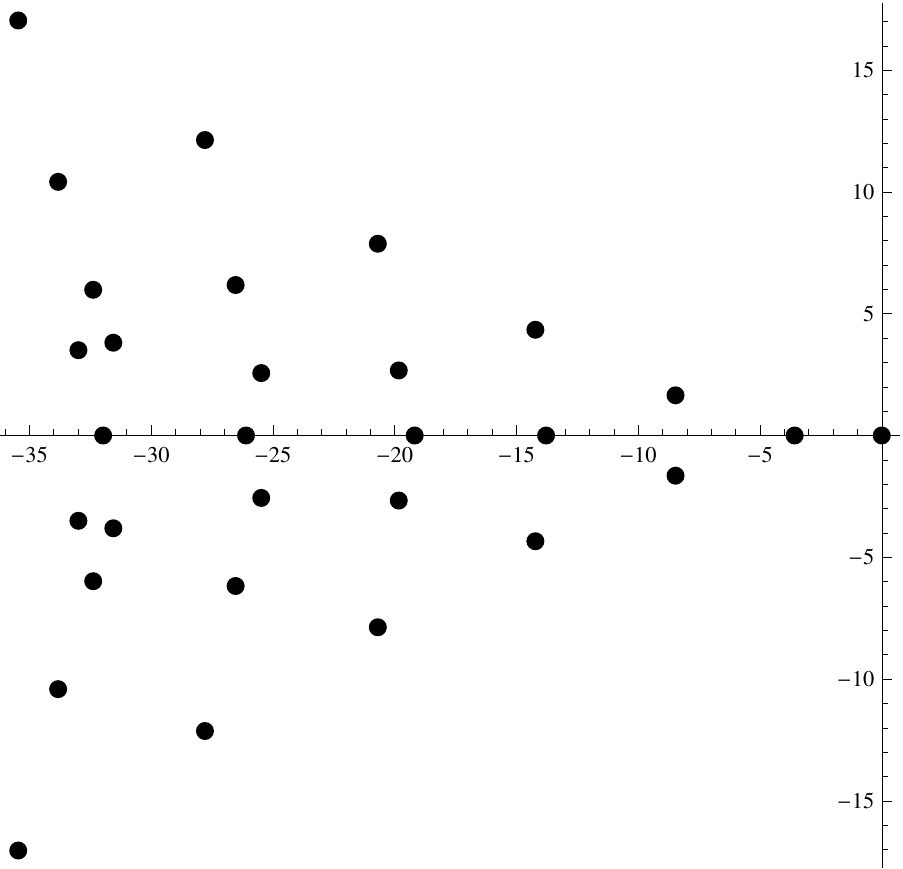}
\end{center}
\caption{First eigenvalues $L^{3/2}E$ of the Markov matrix $M(\gamma=0)$ when $L\to\infty$, computed from (\ref{vp_eq1}), (\ref{vp_eq2}), (\ref{chi_P}).}
\label{fig E}
\end{figure}

In the region $\Re~\gamma\geq0$, all eigenstates in the KPZ regime may be labelled uniquely by a finite set $P$ of non-zero integers with as many positive and negative elements,
\begin{equation}
|P_{+}|=|P_{-}|\;,
\end{equation}
where $P_{+}$ and $P_{-}$ are respectively the positive and negative elements of $P$. The set $P$ is interpreted in section~\ref{section Bethe ansatz} in terms of particle-hole excitations labelling the Bethe eigenstates.

\begin{table}
\begin{center}
$\begin{array}{|c|l|l|}
\hline
P & \qquad v & \qquad \chi_P(v)\\\hline
\emptyset & -\infty & 0  \\\hline
\{-1,1\} & 4.22823 & -3.57806\\\hline
\{-2,1\} & 6.02264\, -0.593574 \rmi & -8.46731-1.64993 \rmi\\
\{-1,2\} & 6.02264\, +0.593574 \rmi & -8.46731+1.64993 \rmi\\\hline
\{-2,2\} & 7.47802 & -13.782\\
\{-3,1\} & 7.28709\, -1.07616 \rmi & -14.2212-4.34068 \rmi\\
\{-1,3\} & 7.28709\, +1.07616 \rmi & -14.2212+4.34068 \rmi\\\hline
\{-2,-1,1,2\} & 8.96685 & -19.1775\\
\{-3,2\} & 8.58003\, -0.505733 \rmi & -19.8304-2.67093 \rmi\\
\{-2,3\} & 8.58003\, +0.505733 \rmi & -19.8304+2.67093 \rmi\\
\{-4,1\} & 8.29833\, -1.49364 \rmi & -20.6905-7.8732 \rmi\\
\{-1,4\} & 8.29833\, +1.49364 \rmi & -20.6905+7.8732 \rmi\\\hline
\{-3,-1,1,2\} & 9.94552\, -0.418264 \rmi & -25.4766-2.56307 \rmi\\
\{-2,-1,1,3\} & 9.94552\, +0.418264 \rmi & -25.4766+2.56307 \rmi\\
\{-3,3\} & 9.59118 & -26.1022\\
\{-4,2\} & 9.48765\, -0.946301 \rmi & -26.5289-6.17958 \rmi\\
\{-2,4\} & 9.48765\, +0.946301 \rmi & -26.5289+6.17958 \rmi\\
\{-5,1\} & 9.15505\, -1.86586 \rmi & -27.788-12.1323 \rmi\\
\{-1,5\} & 9.15505\, +1.86586 \rmi & -27.788+12.1323 \rmi\\\hline
\end{array}$
\end{center}
\caption{Numerical values when $L\to\infty$ of higher gaps $4L^{3/2}E_{P}\simeq\chi_P(v)$, $\chi_P'(v)=0$ for the Markov matrix $M$ with fugacity $\gamma=0$.}
\label{table_eigenstates} 
\end{table}

\end{subsection}

\begin{subsection}{Riemann surface \texorpdfstring{$\mathcal{R}$}{R}}
A meromorphic multivalued function whose various branches are related to one another by analytic continuations is better described as single-valued meromorphic function on a Riemann surface which is a branched covering of the complex plane, see for instance \cite{CM2016.1}. The topology of the Riemann surface is entirely determined by the initial function, by uniqueness of the analytic continuation.

In the case of the multivalued function with branches $\chi_{P}$ described in the previous section, this leads to a non-compact Riemann surface $\mathcal{R}$ of infinite genus built by gluing together all sheets $\mathbb{C}_{P}$ with $P$ a finite subset of $\mathbb{Z}^{*}$ such that $|P_{+}|=|P_{-}|$. The sheets $\mathbb{C}_{P}$ are understood as the domains of definition of the branches $\chi_{P}$. Denoting by $[v,P]$, $v\in\mathbb{C}$ the points of the sheet $\mathbb{C}_{P}$ of $\mathcal{R}$, this leads to a meromorphic function $\mathfrak{X}$ on $\mathcal{R}$ defined by $\mathfrak{X}([v,P])=\chi_{P}(v)$, which is the natural extension to $\mathcal{R}$ of the function $\chi$ defined in (\ref{chi_integral}).

The functions $\chi_{P}$ are holomorphic, except for possible branch points in $2\rmi\pi(\mathbb{Z}+1/2)$. The choice of corresponding branch cuts is largely arbitrary, and determines how $\mathcal{R}$ is partitioned into sheets. We choose in section~\ref{section a.c.} the branch cut $\rmi(-\infty,-\pi]\cup\rmi[\pi,\infty)$. Crossing the cut between $2\rmi\pi(n-1/2)$ and $2\rmi\pi(n+1/2)$, $n\in\mathbb{Z}$ from the sheet $\mathbb{C}_{P}$ of $\mathbb{R}$ either from the left or from the right leads respectively to the sheets $\mathbb{C}_{A_{n}^{\text{l}}P}$ or $\mathbb{C}_{A_{n}^{\text{r}}P}$, where the operators $A^{\text{l}}$, $A^{\text{r}}$ are defined in (\ref{Al P}), (\ref{Ar P}). The two sheets $\mathbb{C}_{A_{n}^{\text{l}}P}$ and $\mathbb{C}_{A_{n}^{\text{r}}P}$ may be distinct unlike for the corresponding Riemann surface for the KPZ fixed point with periodic boundaries \cite{P2020.1} built from half-integer polylogarithms.
\end{subsection}

\end{section}

\begin{section}{Analytic continuations from stationary large deviations of the current}
\label{section a.c.}
In this section, we obtain explicit expressions for analytic continuations of the function $\chi$ in terms of which large deviations of the current in the stationary state are expressed.

\begin{subsection}{Translations and analytic continuations}
\label{section T A}
We consider functions analytic in the domain
\begin{equation}
\label{D}
\mathbb{D}=\mathbb{C}\setminus(\rmi(-\infty,-\pi]\cup\rmi[\pi,\infty))\;,
\end{equation}
with possible branch points in $2\rmi\pi(\mathbb{Z}+1/2)$, and whose analytic continuation along any finite path avoiding the set $2\rmi\pi(\mathbb{Z}+1/2)$ can be extended to an analytic function on $\mathbb{D}$. Using terminology from resurgent functions, we call such functions $2\rmi\pi(\mathbb{Z}+1/2)$-continuable.

Let $f$ be a $2\rmi\pi(\mathbb{Z}+1/2)$-continuable function. The functions $\mathcal{T}_{\text{l}}^{n}f$ and $\mathcal{T}_{\text{r}}^{n}f$, $n\in\mathbb{Z}$, defined \footnote{The index $\text{l}$ stands for left, the index $\text{r}$ for right. We use the notation $\text{l|r}$ for either $\text{l}$ or $\text{r}$} by $(\mathcal{T}_{\text{l|r}}^{n}f)(v)=f(v+2\rmi\pi n)$ respectively for $\Re~v<0$ and for $\Re~v>0$ and extended to $\mathbb{D}$ by analytic continuation, are also $2\rmi\pi(\mathbb{Z}+1/2)$-continuable. For $n=1$, we write simply $\mathcal{T}_{\text{l|r}}$ instead of $\mathcal{T}_{\text{l|r}}^{1}$.

We also define from $f$ $2\rmi\pi(\mathbb{Z}+1/2)$-continuable the functions $\mathcal{A}_{n}^{\text{l}}f$ and $\mathcal{A}_{n}^{\text{r}}f$obtained from $f$ by analytic continuation across the cut $2\rmi\pi(n-1/2,n+1/2)$, respectively from the left and from the right. By construction, the functions $\mathcal{A}_{n}^{\text{l|r}}f$ are also $2\rmi\pi(\mathbb{Z}+1/2)$-continuable.

Translation and analytic continuation operators verify the algebra $\mathcal{A}_{m+n}^{\text{l}}=\mathcal{T}_{\text{r}}^{-n}\mathcal{A}_{m}^{\text{l}}\mathcal{T}_{\text{l}}^{n}$, $\mathcal{A}_{m+n}^{\text{r}}=\mathcal{T}_{\text{l}}^{-n}\mathcal{A}_{m}^{\text{r}}\mathcal{T}_{\text{r}}^{n}$. Since $2\rmi\pi(\mathbb{Z}+1/2)$-continuable functions are analytic on $\mathbb{D}$, $\mathcal{A}_{0}^{\text{l|r}}$ is the identity operator, and one has
\begin{eqnarray}
\label{A[T]}
&& \mathcal{A}_{n}^{\text{l}}=\mathcal{T}_{\text{r}}^{-n}\mathcal{T}_{\text{l}}^{n}\\
&& \mathcal{A}_{n}^{\text{r}}=\mathcal{T}_{\text{l}}^{-n}\mathcal{T}_{\text{r}}^{n}\;,\nonumber
\end{eqnarray}
from which analytic continuations follow easily from translations by $2\rmi\pi$.
\end{subsection}

\begin{subsection}{Lambert functions}
In preparation for the analytic continuation of the function $\chi$ from (\ref{chi_integral}), we study analytic continuations of Lambert functions $W_j$, solutions of $W_j(z)\,\rme^{W_j(z)}=z$ such that $W_j(z)+\log W_j(z)=\log z + 2 \rmi \pi j$, with $\log$ the complex logarithm with branch cut $\mathbb{R}^-$.

We introduce $w_j$, $j\in\mathbb{Z}$ analytic on $\mathbb{D}$ and equal to $W_j(e^{-v-1})$ for $-\pi < \Im~v < \pi$ by
\begin{equation}
w_0(v)=
\left\{
	\begin{array}{ll}
		W_{-[\frac{\Im~v}{2\pi}]}(\rme^{-1-v}) & \text{if} \quad \Re~v < 0\\[2mm]
		W_0 (\rme^{-1-v}) & \text{if} \quad \Re~v > 0
	\end{array}
\right.
\end{equation}
and for $j\in\mathbb{Z}^{*}$
\begin{equation}
\fl\hspace{2mm}
w_j(v)=\left\{ \begin{array}{ll}
	W_{j-[\frac{\Im~v}{2\pi}]-\sgn (j)}(\rme^{-1-v}) & \text{if} \; \Re~v < 0 \; \text{and} \; \sgn(j)\Im(v) > 2\pi (|j|-\frac{1}{2})\\
	W_{[\frac{\Im~v}{2\pi}]} (\rme^{-1-v}) & \text{otherwise}
\end{array}\right.,
\end{equation}
where the square brackets $[\,\cdot\,]$ denote rounding to the nearest integer. The functions $w_{j}$ verify
\begin{equation}
\mathcal{T}_{\text{l}}w_{j}=w_{j-1}
\end{equation}
and
\begin{equation}
\mathcal{T}_{\text{r}}w_{j}=
\left\{
\begin{array}{ll}
w_{0} & j=0\\
w_{-1} & j=1\\
w_{j-1} & j\notin\{0,1\}
\end{array}
\right.\;,
\end{equation}
from which analytic continuations between the $w_{j}$ follow from (\ref{A[T]}).

We consider next square roots of the functions $w_{j}$, and define the functions $y_{j}$ analytic on $\mathbb{D}$ by
\begin{equation}
\label{yj}
y_j(v)=
\left\{
	\begin{array}{lll}
         \sgn(\Im~v - 2\pi j)\sqrt{-w_j(v)} && \Re~v < 0\\
         (-1)^{-[\frac{\Im~v}{2\pi}]} \sqrt{-w_0(v)} && \Re~v > 0 \quad \text{and} \quad j=0\\
         -\sgn (j)\sqrt{-w_j(v)} && \Re~v > 0 \quad \text{and} \quad j \neq 0
	\end{array}
\right.\;,
\end{equation}
which verify
\begin{equation}
\label{Tl y}
\mathcal{T}_{\text{l}}y_{j}=y_{j-1}
\end{equation}
and
\begin{equation}
\label{Tr y}
\mathcal{T}_{\text{r}}y_{j}=
\left\{
\begin{array}{ll}
-y_{0} & j=0\\
-y_{-1} & j=1\\
y_{j-1} & j\notin\{0,1\}
\end{array}
\right.\;.
\end{equation}
From (\ref{A[T]}), analytic continuations between the functions $\pm y_{j}$ is given by
\begin{equation}
\label{Al y}
\mathcal{A}_{n}^{\text{l}} y_{j}
=\left\{
	\begin{array}{ll}
		(-1)^n y_0 & \text{if} \quad j=n\\
		-y_{j+\sgn (n)} & \text{if} \quad j \in \{0\} \cup B_{n-\sgn (n)}\\
		y_j & \text{if} \quad j \notin \{0\} \cup B_{n}
	\end{array}
\right.
\end{equation}
and
\begin{equation}
\label{Ar y}
\mathcal{A}_{n}^{\text{r}} y_j
=\left\{
	\begin{array}{ll}
		(-1)^n y_n \quad & \text{if} \quad j=0\\
		-y_{j-\sgn (n)} \quad & \text{if} \quad j \in B_{n}\\
		y_j \quad & \text{if} \quad j \notin \{0\} \cup B_{n}
	\end{array}
\right.\;,
\end{equation}
where
\begin{equation}
\label{B}
B_n = \left\{
\begin{array}{cl}
	\{1,\ldots,n\} \quad & \text{if} \quad n > 0\\
	\emptyset \quad & \text{if} \quad n = 0\\
	\{n,\ldots,-1\} \quad & \text{if} \quad n < 0
\end{array}
\right.\;.
\end{equation}
We observe that $(\mathcal{A}_{n+1}^{\text{r}}\mathcal{A}_{n}^{\text{l}})^{2}\,y_j=(\mathcal{A}_{n+1}^{\text{l}}\mathcal{A}_{n}^{\text{r}})^{2}\,y_j=y_j$ for any $j,n\in\mathbb{Z}$. Considering small loops with winding number $2$ around $2\rmi\pi(n+1/2)$, this implies that all the branch points of the functions $y_{j}$ are of square root type.
\end{subsection}

\begin{subsection}{Function \texorpdfstring{$\chi_{\emptyset}$}{chi\_0}}
The function $\chi$ defined in (\ref{chi_integral}) has horizontal branch cuts corresponding to poles of the integrand, see figure~\ref{analytic_continuation_domains}. In order to exploit (\ref{A[T]}) for analytic continuations, it will be more convenient to define instead a function $\chi_{\emptyset}$ with vertical branch cuts analytic in the domain $\mathbb{D}$ by rotating the cuts of $\chi$ of $\pm\pi/2$ to the imaginary axis, leaving the function analytic across the segment $(-\rmi \pi,\rmi \pi)$.

\begin{figure}
\begin{center}
\includegraphics{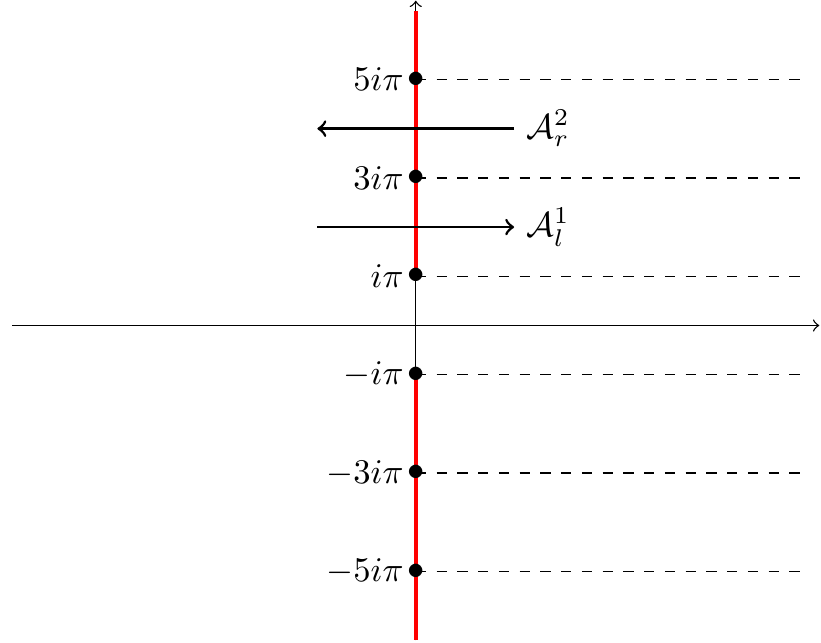}
\end{center}
\caption{Branch cuts of the functions $\chi$ (dashed lines) defined in (\ref{chi_integral}), and $\chi_{\emptyset}$ (solid, red lines) defined in (\ref{chi0}) and analytic in the domain $\mathbb{D}$. Black dots indicate the branch points of $\chi$ and $\chi_\emptyset$.
} 
\label{analytic_continuation_domains}
\end{figure}

The branch cuts of $\chi(v)$ are the half-lines in the complex plane for $v$ on which the integrand 
\begin{equation}
\label{integrand_chi}
f_v(y)=\frac{(1-y^2)(3-y^2)}{1+y^{-2}\,\rme^{y^2-v-1}}
\end{equation}
in (\ref{chi_integral}) has real poles in the variable $y$. Since $0 \leq y^{2}e^{1-y^2} \leq 1$ for $y \in \mathbb{R}$, we find that the branch cuts of $\chi$ are thus the half-lines $\mathbb{R}^{+}+2\rmi\pi a$, $a\in\mathbb{Z}+1/2$. Poles of $f_v$ may be expressed in terms of the square roots of Lambert functions $\pm y_j$ defined in the previous section. More precisely, the branch cut $\mathbb{R}^{+}+2\rmi\pi(j+1/2)$ corresponds to the four poles $\pm y_0(v)$, $\pm y_j(v)$, which take real values on the cut.

\begin{figure}
\begin{center}
\includegraphics{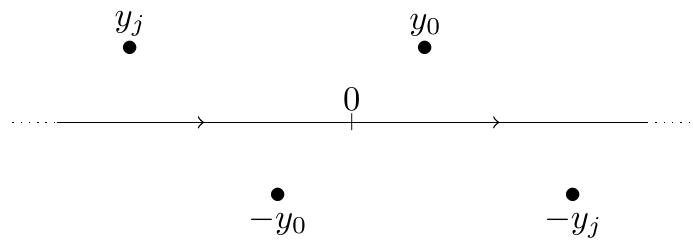}

\vspace{0.5cm}

\includegraphics{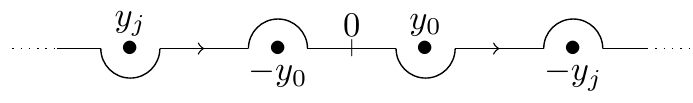}

\vspace{0.5cm}

\includegraphics{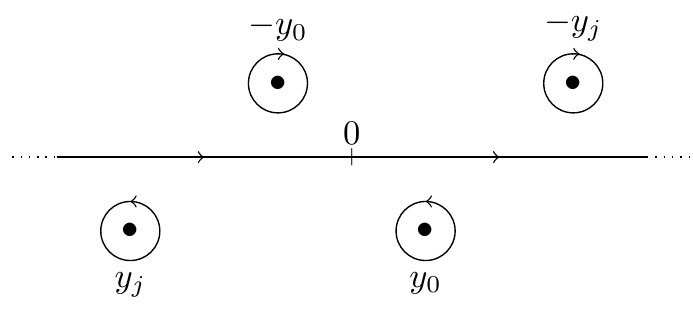}
\end{center}
\caption{Deformation of the integration path in (\ref{chi_integral}) corresponding to analytic continuation of $\chi(v)$ across the cut $\mathbb{R}^{+}+2\rmi\pi(j-1/2)$ when increasing $\Im~v$. The dots represent the corresponding poles of $f_v(y)$ at the cut.}
\label{residu_chi}
\end{figure}

In order to define from $\chi$ a function $\chi_{\emptyset}$ analytic in $\mathbb{D}$, one has to perform the analytic continuation of $\chi$ on the paths $v\pm\rmi t$ with $t$ increasing from $0$ and $v>0$. In the integral expression (\ref{chi_integral}), the analytic continuation simply corresponds to deforming the contour of integration in order to avoid the poles. Alternatively, one can keep the path of integration $\mathbb{R}$ like in (\ref{chi_integral}) and add the residues of the corresponding poles, see figure~\ref{residu_chi}. In particular, when crossing the cut $\mathbb{R}^{+}+2 \rmi \pi(j+\frac{1}{2})$, $j\in\mathbb{Z}$, one finds
\begin{eqnarray}
\lim_{\epsilon \rightarrow 0} \chi(v + \rmi \epsilon) - \chi(v-\rmi \epsilon) = 2 \rmi \pi \Big( \mathrm{res}(f_v,y_j(v)) - \mathrm{res}(f_v,-y_0(v))\\
\hspace{60mm} + \mathrm{res}(f_v,y_0(v)) - \mathrm{res}(f_v,-y_j(v)) \Big)\;. \nonumber
\end{eqnarray}
Computing the residues, the function $\chi_\emptyset$ we seek is defined for $\Re~v > 0$ as
\begin{equation}
\label{chi0}
\chi_\emptyset (v) = \chi(v) - 2 i \Bigg(\frac{1-(-1)^{[\Im\frac{v}{2 \pi}]}}{2} \, \eta_0 (v) + \sum_{j \in B_{[\Im\frac{v}{2\pi}]}} \eta_j (v) \Bigg) \;,
\end{equation}
with the set $B_n$ defined in (\ref{B}) and
\begin{equation}
\label{etaj}
\eta_j(v) = \frac{y_j(v)^3}{3} - y_j(v)\;.
\end{equation}
The functions $\eta_j$ verify $\eta_{j}'(v)=y_{j}(v)/2$. For $\Re~v < 0$, $\chi_\emptyset (v)=\chi(v)$, and $\chi_\emptyset$ is analytic in $\mathbb{D}$. The difference between $\chi_\emptyset$ and $\chi$ as defined by (\ref{chi_integral}) is merely a different choice of convention for the branch cuts.
\end{subsection}

\begin{subsection}{Analytic continuation of \texorpdfstring{$\chi_{\emptyset}$}{chi\_0}: functions \texorpdfstring{$\chi_{P}$}{chi\_P}}
We consider now analytic continuations of $\chi_{\emptyset}$ defined in $\mathbb{D}$ across the cuts $(2\rmi\pi(n-1/2),2\rmi\pi(n-1/2))$ using the notations from section~\ref{section T A}.

Since $\chi$ defined in (\ref{chi_integral}) is analytic in all the strips $\{v\in\mathbb{C},2\rmi\pi(n-1/2)<\Im~v<2\rmi\pi(n-1/2)\}$, one has $(\mathcal{A}_{n}^{\text{l}}\chi_{\emptyset})(v)=\chi(v)$ for $\Re\,v>0$. From the relation (\ref{chi0}) between $\chi$ and $\chi_{\emptyset}$, this leads to
\begin{equation}
\label{Al chi0}
\mathcal{A}_{n}^{\text{l}} \chi_\emptyset = \chi_\emptyset + 2i \Bigg(\frac{1-(-1)^n}{2} \eta_0 + \sum_{j \in B_n} \eta_j \Bigg)\;,
\end{equation}
which is an identity between functions defined in $\mathbb{D}$.

Similarly, for the analytic continuation from the right, $(\mathcal{A}_{n}^{\text{r}}\chi_{\emptyset})(v)$ for $\Re~v<0$ is equal to $\mathcal{A}_{n}^{\text{r}}$ applied to the function of $v$ in the right side of (\ref{chi0}). Using (\ref{Ar y}), this leads to
\begin{equation}
\label{Ar chi0}
\mathcal{A}_{n}^{\text{r}} \chi_\emptyset = \chi_\emptyset + 2i \Bigg(\eta_0 - \frac{1+(-1)^n}{2} \eta_n + \sum_{j \in B_n} \eta_j \Bigg)\;.
\end{equation}

Using (\ref{Al chi0}), (\ref{Ar chi0}) and (\ref{Al y}), (\ref{Ar y}), analytic continuations of $\chi_{\emptyset}$ on arbitrary paths avoiding $2\rmi\pi(\mathbb{Z}+1/2)$ can be obtained by applying operators $\mathcal{A}_{n}^{\text{l|r}}$ whenever the path crosses the cut $(2\rmi\pi(n-1/2),2\rmi\pi(n-1/2))$. We observe that we always obtain functions of the form
\begin{equation}
\chi_P (v) = \chi_\emptyset(v) + 2 \rmi \sum_{j \in P} \eta_j (v) \;,
\end{equation}
with $P$ a finite set of integers.

Analytic continuations of the functions $\chi_{P}$ can be performed in the same way. It is however convenient to consider first translations by integer multiples of $2\rmi\pi$. Writing $\mathcal{T}_{\text{l|r}}^{n}\chi_P = \chi_{T_{\text{l|r}}^{n}P}$, since $\chi_{\emptyset}(v)$ is periodic in $v$ with period $2\rmi\pi$ when $\Re~v<0$, one has from (\ref{Tl y})
\begin{equation}
T_{\text{l}}^{-n}P=P+n\;.
\end{equation}
After tedious calculations, we also obtain
\begin{equation}
\label{Tr P}
\fl\hspace{2mm}
T_{\text{r}}^{-n}P=
\left\{
	\begin{array}{ll}
		((P+n) \setminus C_{n}) \cup (B_{n} \setminus (P+n+\sgn~n)) \cup \{0\}
		&
		\begin{array}{ll}
			0\in P \;\&\; n\;\text{even}\\
			0\notin P \;\&\; n\;\text{odd}
		\end{array}\\[5mm]
		((P+n) \setminus C_{n}) \cup (B_{n} \setminus (P+n+\sgn~n))
		&
		\begin{array}{ll}
			0\in P \;\&\; n\;\text{odd}\\
			0\notin P \;\&\; n\;\text{even}
		\end{array}
	\end{array}
\right.\!\!,
\end{equation}
with $C_{n}=B_{n}\cup\{0\}$ and the convention $\sgn~0=0$. Analytic continuations of the functions $\chi_{P}$ then follow easily using (\ref{A[T]}). Writing $\mathcal{A}_{n}^{\text{l|r}}\chi_P = \chi_{A_{n}^{\text{l|r}}P}$, one has
\begin{equation}
\label{Al P}
\fl\hspace{10mm}
A_{n}^{\text{l}}P=
\left\{
	\begin{array}{ll}
		(P \setminus C_{n}) \cup (B_{n} \setminus (P+\sgn~n)) \cup \{0\}
		&
		\begin{array}{ll}
			n\in P \;\&\; n\;\text{even}\\
			n\notin P \;\&\; n\;\text{odd}
		\end{array}\\[5mm]
		(P \setminus C_{n}) \cup (B_{n} \setminus (P+\sgn~n))
		&
		\begin{array}{ll}
			n\in P \;\&\; n\;\text{odd}\\
			n\notin P \;\&\; n\;\text{even}
		\end{array}
	\end{array}
\right.
\end{equation}
while $A_{0}^{\text{r}}P=P$ and for $n\neq0$
\begin{equation}
\label{Ar P}
\fl\hspace{10mm}
A_{n}^{\text{r}}P=
\left\{
	\begin{array}{ll}
		(P \setminus C_{n}) \cup (C_{n} \setminus (P-\sgn~n)) \cup \{n\}
		&
		\begin{array}{ll}
			0\in P \;\&\; n\;\text{even}\\
			0\notin P \;\&\; n\;\text{odd}
		\end{array}\\[5mm]
		(P \setminus C_{n}) \cup (C_{n-\sgn~n} \setminus (P-\sgn~n))
		&
		\begin{array}{ll}
			0\in P \;\&\; n\;\text{odd}\\
			0\notin P \;\&\; n\;\text{even}
		\end{array}
	\end{array}
\right.\!\!.
\end{equation}
We observe in particular that $(A_{n+1}^{\text{r}}A_{n}^{\text{l}})^{2}\,P=(A_{n+1}^{\text{l}}A_{n}^{\text{r}})^{2}\,P=P$ for any $P\subset\mathbb{Z}$ and $n\in\mathbb{Z}$, which implies that all the branch points of the functions $\chi_{P}$ are of square root type.

The operator $T_{\text{r}}$ generates a group acting on sets of integers. When computing $e=\chi_{P}(v)$ for $v$ solution of the equation $\chi_{P}'(v)=s$, which is needed for the calculation of the eigenvalues of open TASEP in the KPZ regime, all the sets $P$ in the same orbit under the action of this group give the same value of $e$ if $\Re~v>0$. Thus, eigenvalues must not be indexed by sets $P$ but rather by equivalence classes of sets in the same orbit. We observe that for any finite set $P\subset\mathbb{Z}$, there always exists a single element $P^{*}$ of the orbit of $P$ under the action of $T_{\text{r}}$ which has the same number of positive and negative elements, see \ref{appendix Tr}. Furthermore, when restricting to sets $P$ such that $\chi_{P}$ is reachable from $\chi_{\emptyset}$ by analytic continuations, we also observe that $P^{*}$ does not contain $0$, see also \ref{appendix Tr}. Thus, assuming that eigenvalues of TASEP of the form (\ref{vp_eq1}), (\ref{vp_eq2}) may be obtained from each other by analytic continuations, the restriction $|P_{+}|=|P_{-}|$, $P\subset\mathbb{Z}^{*}$ comes indeed in a rather natural way.
\end{subsection}

\end{section}

\begin{section}{Direct Bethe ansatz derivation} \label{bethe_ansatz_derivation}
\label{section Bethe ansatz}
In this section we derive the exact asymptotic expression (\ref{vp_eq1}), (\ref{vp_eq2}) for the spectral gaps using Bethe ansatz, starting with the Bethe equations obtained by Cramp\'e and Nepomechie in \cite{CN2018.1}. We first solve the Bethe equations for finite system sizes and find a parametrization of the relevant solutions as particle-hole excitations over a Fermi sea, before computing the large $L$ asymptotics.

\begin{subsection}{Bethe equations}
Setting
\begin{equation}
a=\frac{1}{\alpha}-1
\qquad\text{and}\qquad
b=\frac{1}{\beta} -1\;,
\end{equation}
every eigenvalue of $M(\gamma)$ can be expressed \cite{CN2018.1} in terms of $L+2$ Bethe roots $u_j$, $0 \leq j \leq L+1$ satisfying the Bethe equations
\begin{equation}
\fl\hspace{5mm} u_j^L(u_j+a)(u_j+b)(au_j+1)(bu_j+1)=(-1)^{L+1} \rme^{2\gamma}(1-u_j)^{2L+2}(u_j+1)^2 \prod_{k=0}^{L+1} u_k \;.
\end{equation}
The eigenvalue of $M(\gamma)$ corresponding to a set $\{u_0,\ldots,u_{L+1}\}$ of Bethe roots is $E = -\Lambda ' (1) /2$, where $\Lambda(x)$ is the corresponding eigenvalue of the commuting family of transfer matrices with spectral parameter $x$, given in terms of Baxter's $Q$ function $Q(x)=\prod_{k=0}^{L+1} (x-u_k)$ by \cite{CN2018.1}
\begin{equation}
\label{Lambda_vp}
\Lambda (x)= \frac{x^{L}(x+a)(x+b) }{\rme^{\gamma}\,Q(x)} + \rme^\gamma \, \frac{Q(0)}{Q(x)} \frac{(1-x)^{2L+2} (x+1)^2}{(ax+1)(bx+1)} \;.
\end{equation}
More explicitly, $\Lambda(1)=1$, obtained in \cite {CN2018.1} from another set of Bethe equations, implies
\begin{equation}
\label{product_bethe_roots}
Q(1)=\frac{\rme^{-\gamma}}{\alpha \beta}\;,
\end{equation}
and one has
\begin{equation}
\label{E[uj]}
E = 1 - \frac{\alpha + \beta}{2} + \frac{1}{2} \sum_{j=0}^{L+1} \frac{u_j}{1-u_j} \;.
\end{equation}
\end{subsection}

\begin{subsection}{Functions \texorpdfstring{$\mathfrak{u}_{j}(C)$}{u\_j(C)} and \texorpdfstring{$\mathfrak{u}_{j}^{*}(C)$}{u\_j*(C)}}
Introducing the parameter $C=-\rme^{2\gamma}\prod_{k=0}^{L+1} u_k$ conjugate to $\gamma$, the $L+2$ Bethe roots $u_j$, $j=0,\ldots,L+1$ are among the $2L+4$ solutions $u$ of the polynomial equation $R(u,C)=0$ with
\begin{equation}
\fl\hspace{10mm}
R(u,C)=u^L(u+a)(u+b)(au+1)(bu+1)-(-1)^{L} C(1-u)^{2L+2}(u+1)^2\;,
\end{equation}
which belong to the curve
\begin{equation}
\label{Gamma C}
\Gamma_{C}:|u|^L|u+a|\,|u+b|\,|au+1|\,|bu+1|=|C|\,|1-u|^{2L+2}|u+1|^2\;,
\end{equation}
plotted for the special case $\alpha=\beta=1$ in figure~\ref{fig curve C}. The curve $\Gamma_{C}$ is the analogue of the generalized Cassini ovals that are found instead for periodic TASEP \cite{GM2004.1}.

\begin{figure}
\begin{center}
\begin{tabular}{lll}
	\begin{tabular}{l}\includegraphics[width=45mm]{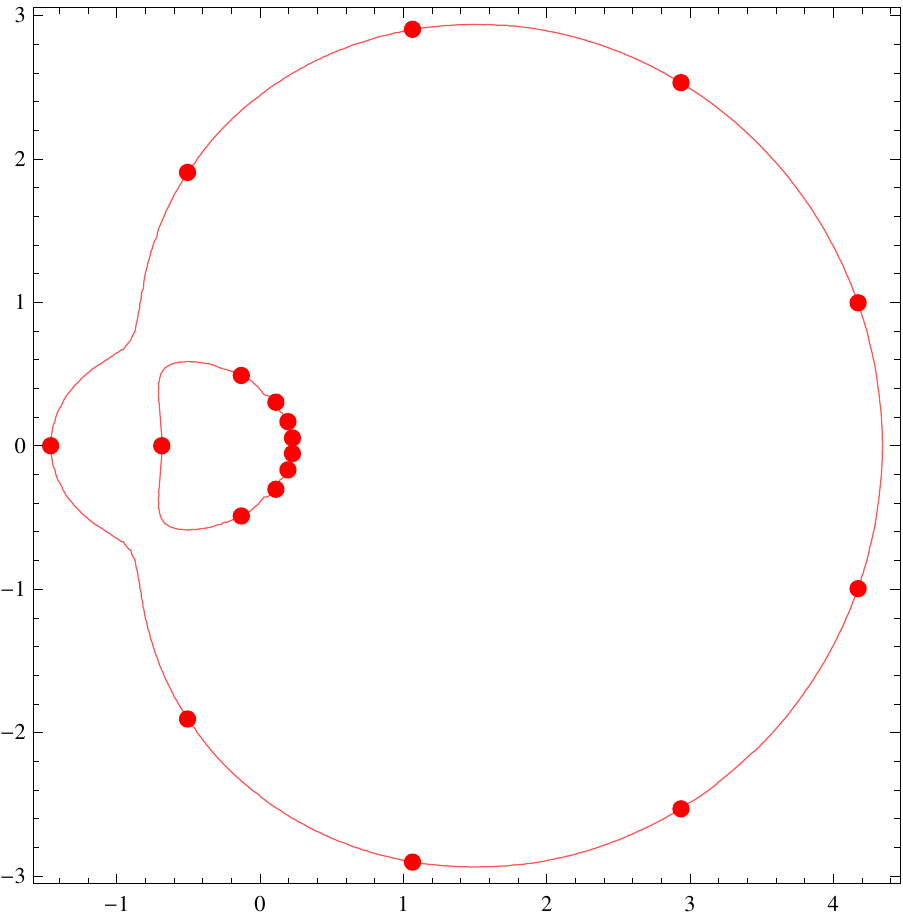}\end{tabular}
	&
	\begin{tabular}{l}\includegraphics[width=45mm]{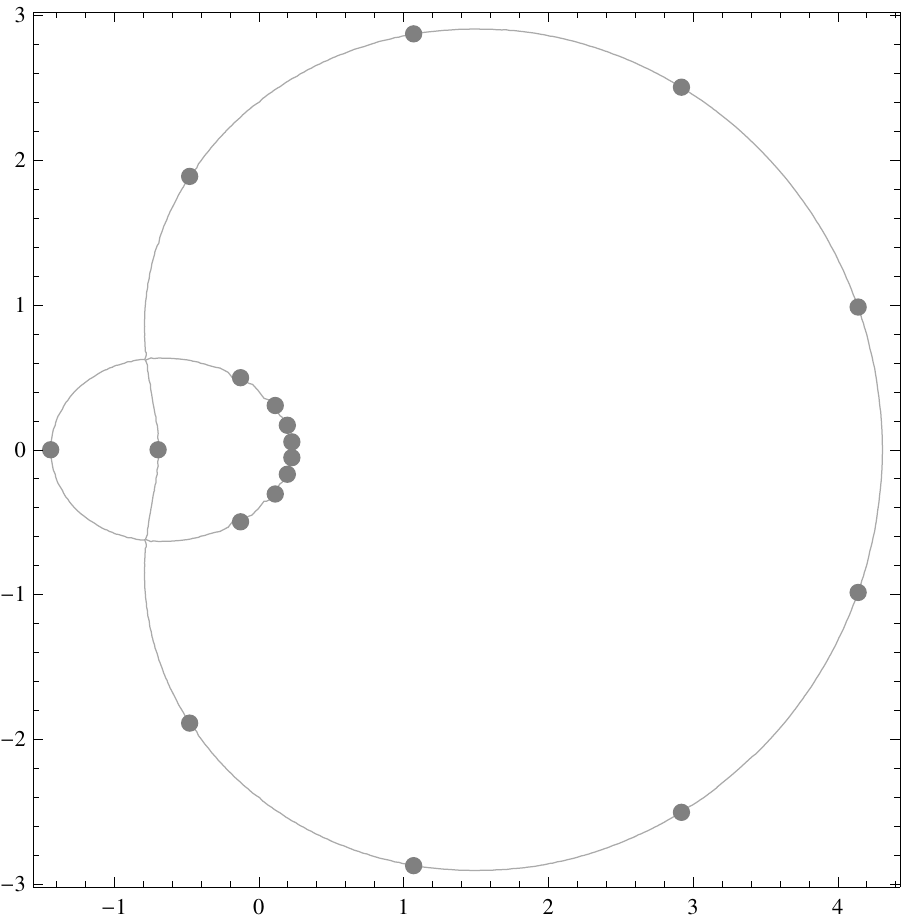}\end{tabular}
	&
	\begin{tabular}{l}\includegraphics[width=45mm]{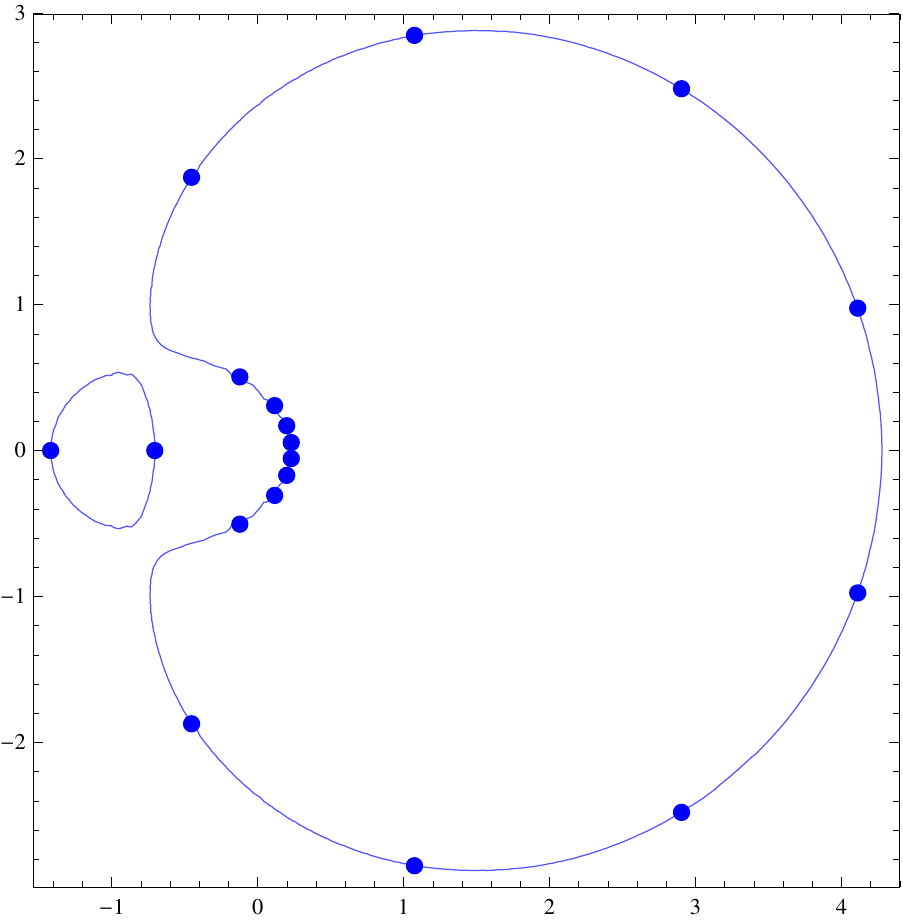}\end{tabular}
\end{tabular}
\end{center}
\caption{Curve $\Gamma_{C}$ defined in (\ref{Gamma C}), plotted for $L=7$, $\alpha=\beta=1$, with $C=0.9\,C^{*}$ (left), $C=C^{*}$ (middle) and $C=1.1\,C^{*}$ (right).}
\label{fig curve C}
\end{figure}

The curve $\Gamma_{C}$ has a double point at $u$ whenever both equations $R(u,C)=0$ and $\partial_{u}R(u,C)=0$ are satisfied. Solutions $(u,C)$ with finite non-zero $C$ verify
\begin{equation}
\label{eq u crit}
\frac{L}{u}+\frac{1}{u+a}+\frac{1}{u+b}+\frac{1}{u+a^{-1}}+\frac{1}{u+b^{-1}}=\frac{2L+2}{u-1}+\frac{2}{u+1}\;.
\end{equation}
In the special case $\alpha=\beta=1$, the equation (\ref{eq u crit}) has the two solutions $u=(-L\pm2\rmi\sqrt{L+1})/(L+2)$, corresponding to the same value $C=C_{*}$ with
\begin{equation}
\label{C* ab0}
C_{*}=-\frac{(L+2)^{L+2}}{4^{L+2}(L+1)^{L+1}}
\quad\text{when}\;\alpha=\beta=1\;.
\end{equation}
For generic values of $\alpha$ and $\beta$, equation (\ref{eq u crit}) instead has six solutions $u$ corresponding to three values of $C$. In the limit $\alpha,\beta\to1$, two of them converge to zero while the last one, which we still call $C_{*}$, converges to (\ref{C* ab0}), and has for arbitrary values of $a$ and $b$ the large $L$ behaviour
\begin{equation}
C_{*}\simeq-\frac{\rme\,(1-a)^{2}(1-b)^{2}}{4^{L+2}}\Big(L+\frac{3}{2}+\frac{4a}{(1-a)^{2}}+\frac{4b}{(1-b)^{2}}\Big)\;.
\end{equation}
The two extra critical values of $C$ do not contribute to the KPZ regime within the maximal current phase, so that the labelling of Bethe roots introduced below in the case $\alpha=\beta=1$ essentially still applies in the whole maximal current phase, at least for the purpose of computing the spectrum.

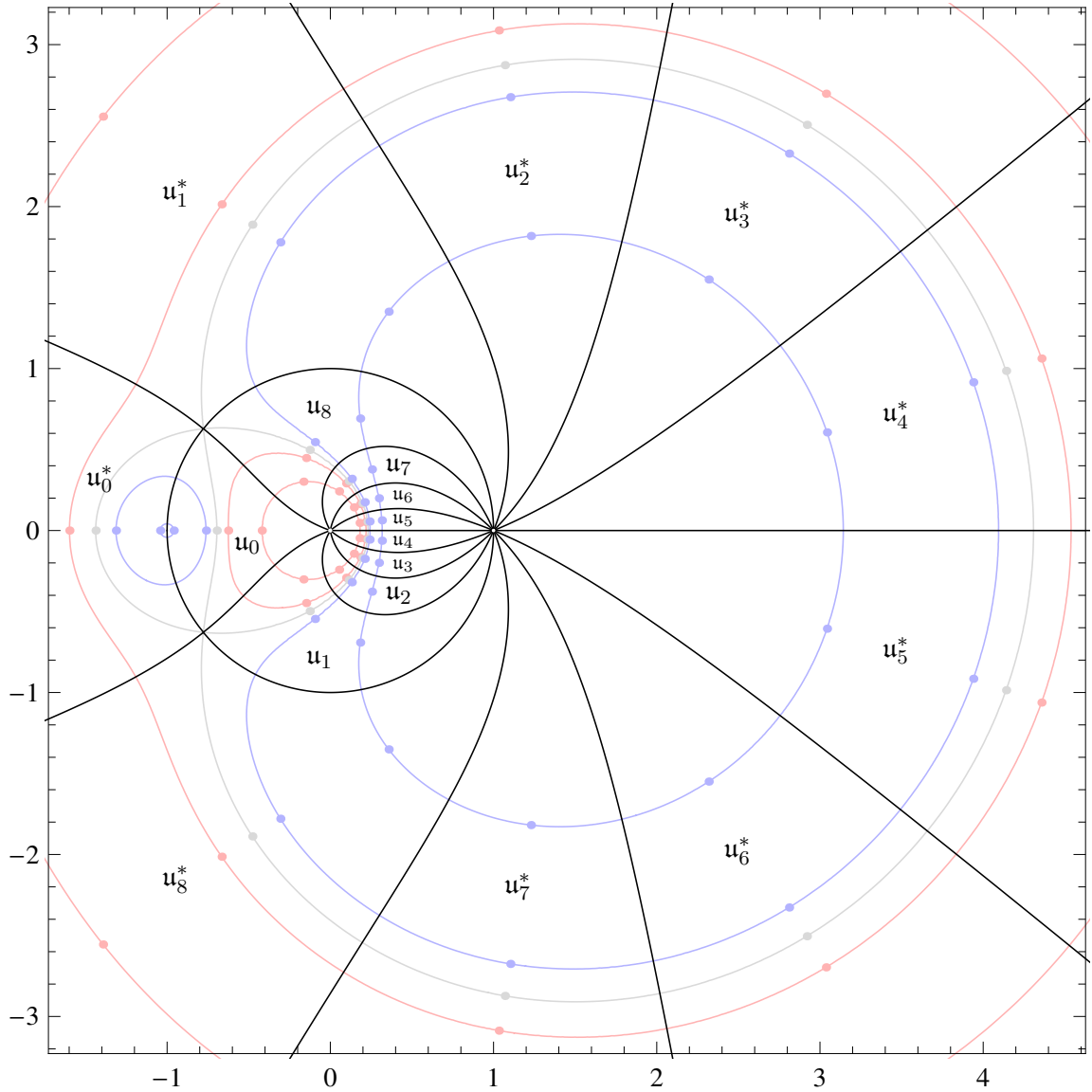
\begin{figure}
\begin{center}
	\includegraphics[width=150mm]{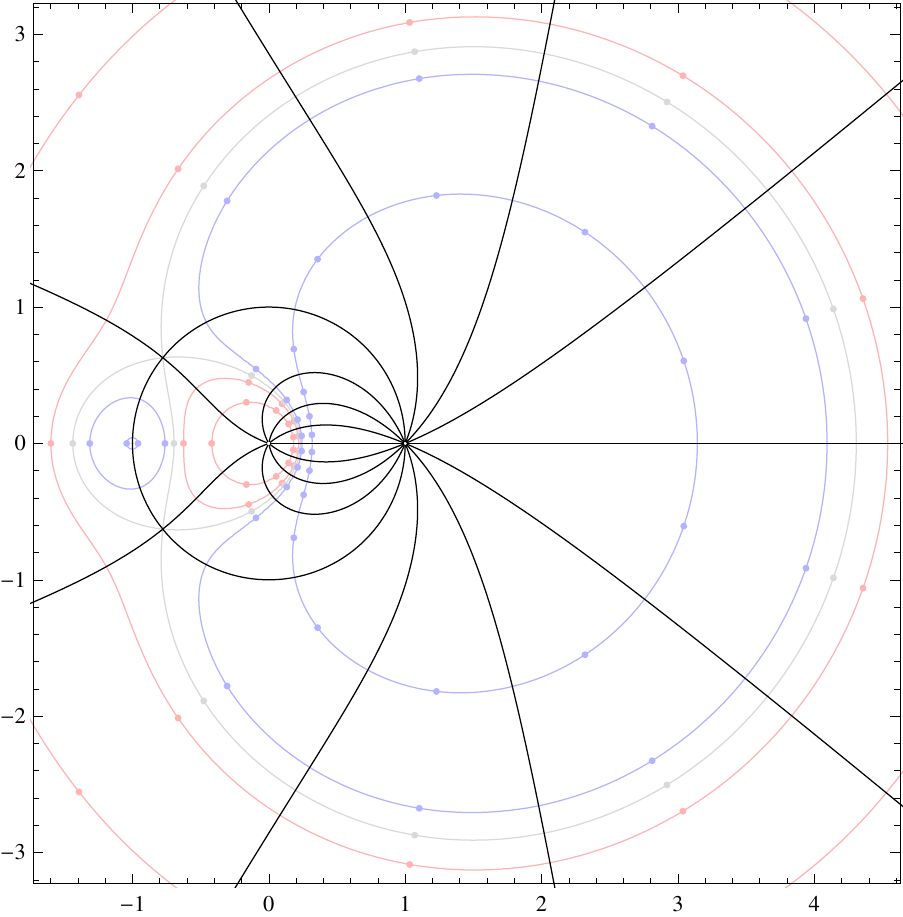}
	\begin{picture}(0,0)
		\put(-120,76){\small$\mathfrak{u}_{0}$}
		\put(-110,60){\small$\mathfrak{u}_{1}$}
		\put(-99,69){\small$\mathfrak{u}_{2}$}
		\put(-98,73.3){\scriptsize$\mathfrak{u}_{3}$}
		\put(-98,76.6){\scriptsize$\mathfrak{u}_{4}$}
		\put(-98,79.7){\scriptsize$\mathfrak{u}_{5}$}
		\put(-98,83){\scriptsize$\mathfrak{u}_{6}$}
		\put(-99,87){\small$\mathfrak{u}_{7}$}
		\put(-110,95){\small$\mathfrak{u}_{8}$}
		\put(-140.5,85){\small$\mathfrak{u}_{0}^{*}$}
		\put(-130,125){\small$\mathfrak{u}_{1}^{*}$}
		\put(-82.5,128){\small$\mathfrak{u}_{2}^{*}$}
		\put(-52,122){\small$\mathfrak{u}_{3}^{*}$}
		\put(-30,94){\small$\mathfrak{u}_{4}^{*}$}
		\put(-30,61){\small$\mathfrak{u}_{5}^{*}$}
		\put(-52,33){\small$\mathfrak{u}_{6}^{*}$}
		\put(-82.5,28){\small$\mathfrak{u}_{7}^{*}$}
		\put(-130,29){\small$\mathfrak{u}_{8}^{*}$}
	\end{picture}
\end{center}
\caption{Labelling of the $2L+2$ solutions $u$ of $R(u,C)=0$ in terms of functions $\mathfrak{u}_{j}(C)$ and $\mathfrak{u}_{j}^{*}(C)$, $j=0,\ldots,L+1$ for $L=7$ and $\alpha=\beta=1$. The domains $\mathfrak{u}_{j}(\mathbb{C}\setminus\mathbb{R}^{-})$ and $\mathfrak{u}_{j}^{*}(\mathbb{C}\setminus\mathbb{R}^{-})$ are delimited by the black curves. The lighter, blue, grey and red curves correspond to the curves $\Gamma_{C}$ for $|C/C^{*}|\in\{1/20,1/2,1,2,100\}$, compare with figure~\ref{fig curve C}. The dots are the corresponding solutions of $R(u,C)=0$ with $C>0$.}
\label{fig uj}
\end{figure}

Focusing on the case $\alpha=\beta=1$, we introduce in the following a consistent labelling scheme for the $2L+4$ solutions of $R(u,C)=0$, with well defined functions $\mathfrak{u}_{j}(C)$ and $\mathfrak{u}_{j}^{*}(C)$, $j=0,\ldots,L+1$ (the star does \emph{not} mean complex conjugation), see figure~\ref{fig uj}. Since double points of $\Gamma_{C}$ happen only for $C\in\{\infty,C_{*},0\}$, it is possible to choose all the functions $\mathfrak{u}_{j}(C)$ and $\mathfrak{u}_{j}^{*}(C)$ analytic in $\mathbb{C}\setminus\mathbb{R}^{-}$, with the single branch cut $\mathbb{R}^{-}$. The labelling can then be made unambiguous from e.g. the small $C$ behaviour of the solutions of $R(u,C)=0$. We choose
\begin{eqnarray}
\label{u C->0}
& \mathfrak{u}_{j}(C)\simeq-\rme^{\frac{2\rmi\pi j}{L+2}}\,C^{\frac{1}{L+2}}\\
& \mathfrak{u}_{j}^{*}(C)\simeq-\rme^{-\frac{2\rmi\pi j}{L+2}}\,C^{-\frac{1}{L+2}}\nonumber
\end{eqnarray}
when $C\to0$, the power $\pm\frac{1}{L+2}$ being understood with branch cut $\mathbb{R}^{-}$. The labelling can be extended to more general $\alpha,\beta$ by continuity. In particular, for arbitrary $\alpha,\beta$ in the maximal current phase, so that $a,b<1$, one has the corresponding large $C$ behaviour
\begin{eqnarray}
\label{u C->inf}
& 1+\mathfrak{u}_{0}(C)\simeq\frac{(1-a)(1-b)}{2^{L+1}\sqrt{C}}\nonumber\\
& 1+\mathfrak{u}_{0}^{*}(C)\simeq-\frac{(1-a)(1-b)}{2^{L+1}\sqrt{C}}\\
& 1-\mathfrak{u}_{j}(C)\simeq\frac{\rme^{-\frac{2\rmi\pi(j-1-L/2)}{2L+2}}}{(4\alpha^{2}\beta^{2}C)^{\frac{1}{2L+2}}}
\qquad\text{for}\;j=1,\ldots,L+1\nonumber\\
& 1-\mathfrak{u}_{j}^{*}(C)\simeq\frac{\rme^{-\frac{2\rmi\pi(j+L/2)}{2L+2}}}{(4\alpha^{2}\beta^{2}C)^{\frac{1}{2L+2}}}
\qquad\text{for}\;j=1,\ldots,L+1\;.\nonumber
\end{eqnarray}

The functions $\mathfrak{u}_{j}(C)$, $\mathfrak{u}_{j}^{*}(C)$ have branch points $0$, $C^{*}$ and $-\infty$. Analytic continuations across the cut $(-\infty,C_{*})$ from above are denoted by the operator $\mathcal{A}_{\text{out}}$, while analytic continuations across the cut $(C_{*},0)$ from above are denoted by the operator $\mathcal{A}_{\text{in}}$. One has, see figure~\ref{fig uj},
\begin{equation}
\begin{array}{lll}
\mathcal{A}_{\text{out}}\mathfrak{u}_{j}=\mathfrak{u}_{j+1} && 0\leq j\leq L\\
\mathcal{A}_{\text{out}}\mathfrak{u}_{L+1}=\mathfrak{u}_{0} && j=L+1\\[2mm]
\mathcal{A}_{\text{out}}\mathfrak{u}_{j}^{*}=\mathfrak{u}_{j+1}^{*} && 0\leq j\leq L\\
\mathcal{A}_{\text{out}}\mathfrak{u}_{L+1}^{*}=\mathfrak{u}_{0}^{*} && j=L+1
\end{array}
\end{equation}
and
\begin{equation}
\label{Ain u}
\begin{array}{lll}
\mathcal{A}_{\text{in}}\mathfrak{u}_{0}=\mathfrak{u}_{0}^{*} && j=0\\
\mathcal{A}_{\text{in}}\mathfrak{u}_{0}^{*}=\mathfrak{u}_{0} && j=0\\[2mm]
\mathcal{A}_{\text{in}}\mathfrak{u}_{j}=\mathfrak{u}_{j+1} && 1\leq j\leq L\\
\mathcal{A}_{\text{in}}\mathfrak{u}_{L+1}=\mathfrak{u}_{1}^{*} && j=L+1\\
\mathcal{A}_{\text{in}}\mathfrak{u}_{j}^{*}=\mathfrak{u}_{j+1}^{*} && 1\leq j\leq L\\
\mathcal{A}_{\text{in}}\mathfrak{u}_{L+1}^{*}=\mathfrak{u}_{1} && j=L+1
\end{array}\;.
\end{equation}
\end{subsection}

\begin{subsection}{Identification of the excited states}
In the notations of the previous section, the stationary eigenstate of $M(\gamma)$ for $\gamma>0$ corresponds \cite{CN2018.1} for the $L+2$ Bethe roots to the choice $u_{j}=\mathfrak{u}_{j}(C)$, $j=0,\ldots,L+1$.

Based on numerical computations for small values of $L$, we conjecture that for any given value of $C\in\mathbb{C}\setminus\mathbb{R}^{-}$, all $2^{L+1}$ choices with $u_{j}\in\{\mathfrak{u}_{j}(C),\mathfrak{u}_{j}^{*}(C)\}$ such that an even number of $\mathfrak{u}_{j}^{*}$ is chosen correspond to eigenvalues of $M(\gamma)$. This is consistent with analytic continuation since the collection of all such choices is stable by repeated applications of $\mathcal{A}_{\text{in}}$, $\mathcal{A}_{\text{out}}$. Of course, $M(\gamma)$ has only $2^{L}$ eigenstates, but the choices above for a fixed value of $C$ correspond to distinct values of $\gamma$, and the ``same'' eigenvalue may thus appear several times with distinct values of $\gamma$.

Taking the $2L+2$-th root of the Bethe equation, one has
\begin{equation}
\label{power b e}
\frac{(1-u_{j})(1+u_{j})^{\frac{1}{L+1}}C^{\frac{1}{2L+2}}}{u^{\frac{L}{2L+2}}(u+a)^{\frac{1}{2L+2}}(u+b)^{\frac{1}{2L+2}}(1+au)^{\frac{1}{2L+2}}(1+bu)^{\frac{1}{2L+2}}}=\rme^{-\frac{2\rmi\pi k_{j}}{2L+2}}\;,
\end{equation}
where $k_{j}$, integer if $L$ is even and half-integer if $L$ is odd, is the momentum of the quasi-particle associated to the Bethe root $u_{j}$. The momentum $k_{j}$ depends on the precise choice of the cuts for the fractional powers in (\ref{power b e}). A natural choice consistent with the large $C$ behaviour (\ref{u C->inf}) of the functions $\mathfrak{u}_{j},\mathfrak{u}_{j}^{*}$ is $k_{j}=j-1-L/2$ for Bethe roots $u_{j}=\mathfrak{u}_{j}(C)$ and $k_{j}=j+L/2$ for Bethe roots $u_{j}=\mathfrak{u}_{j}^{*}(C)$, $j=1,\ldots,L+1$. The stationary state then corresponds to the filled Fermi sea $\{k_{j},j=1,\ldots,L+1\}=\{-L/2,\ldots,L/2\}$.

\begin{figure}
\begin{center}
\includegraphics{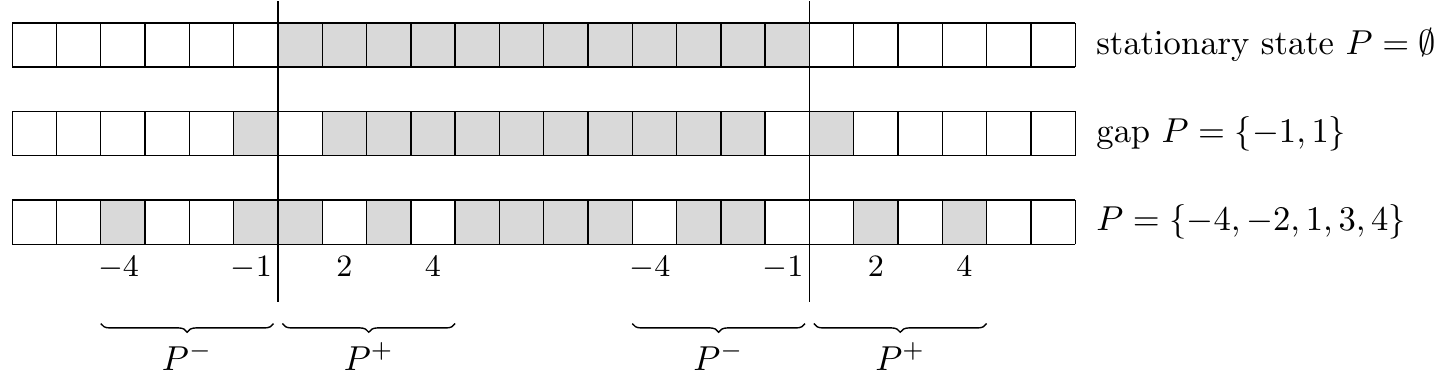}
\end{center}
\caption{Choices of momenta of quasi-particles $k_j$ for some eigenstates contributing to the KPZ regime in the maximal current phase. Filled squares represent values of $-k_j$ for the Bethe roots chosen, and vertical bars correspond to the limits of the interval $\{-L/2,\ldots,L/2\}$.}
\label{fermi_roots}
\end{figure}

Numerics indicate that in the large $L$ limit, only eigenstates corresponding to particle-hole excitations with the same number of excitations at both ends of the Fermi sea contribute to the KPZ regime in the maximal current phase. More precisely, such eigenstates are defined in terms of a finite set $P\subset\mathbb{Z}^{*}$ with the same number of positive and negative elements, $|P_{+}|=|P_{-}|$, and correspond to the choice $u_{0}=\mathfrak{u}_{0}(C)$, $u_{j}=\mathfrak{u}_{j}(C)$ for $j\in\[1,L+1\]\setminus((-P_{-})\cup(L+2-P_{+}))$ and $u_{j}=\mathfrak{u}_{j}^{*}(C)$ for $j\in(-P_{-})\cup(L+2-P_{+})$. This corresponds for the momenta of quasi-particles to $\{-k_{j},j=1,\ldots,L+1\}$ being equal to
\begin{equation}
\fl
\Big(\{-L/2,\ldots,L/2\}\cup(L/2+P_{+})\cup(-L/2+P_{-})\Big)\setminus\Big((-L/2-1+P_{+})\cup(L/2+1+P_{-})\Big),
\end{equation}
see figure~\ref{fermi_roots}, which is consistent with the fact that Umklapp processes exchanging momenta between both sides of the Fermi sea are suppressed in the KPZ regime, as is already known for periodic TASEP, and implies here that the total momentum of quasi-particles is equal to zero,
\begin{equation}
\label{null_sum_kj}
\sum_{k=1}^{L+1} k_j = 0 \quad \mathrm{mod} \quad 2L+2\;.
\end{equation}
\end{subsection}

\begin{subsection}{Large \texorpdfstring{$L$}{L} asymptotics}
We want to show that the eigenvalues corresponding to a given set $P\subset\mathbb{Z}^{*}$, $|P_{+}|=|P_{-}|$ for increasing values of the system size $L$ have the large $L$ asymptotics (\ref{vp_eq1}), (\ref{vp_eq2}). In order to do this we fix the parameter $C$ and split the sum in (\ref{E[uj]}) into the contribution of the functions $\mathfrak{u}_{j}(C)$, corresponding to the stationary state, plus the contribution of particle-hole excitations, as
\begin{eqnarray}
\label{sum_E_asymptotic}
\fl\hspace{10mm}
E = 1-\frac{\alpha+\beta}{2} + \frac{1}{2}\sum_{j=0}^{L+1}\frac{\mathfrak{u}_{j}(C)}{1-\mathfrak{u}_{j}(C)}
+ \frac{1}{2}\sum_{j \in L+2-P^{+}} \Big(\frac{\mathfrak{u}_{j}^{*}(C)}{1-\mathfrak{u}_{j}^{*}(C)} - \frac{\mathfrak{u}_{j}(C)}{1-\mathfrak{u}_{j}(C)}\Big)\\
\hspace{53mm}
+ \frac{1}{2}\sum_{j \in -P^{-}} \Big(\frac{\mathfrak{u}_{j}^{*}(C)}{1-\mathfrak{u}_{j}^{*}(C)} - \frac{\mathfrak{u}_{j}(C)}{1-\mathfrak{u}_{j}(C)}\Big)\;.\nonumber
\end{eqnarray}
The parameter $C$ is then fixed in terms of the fugacity $\gamma$ from (\ref{product_bethe_roots}), which is likewise split as
\begin{eqnarray}
\label{sum_gamma_asymptotic}
\gamma = - \log (\alpha\beta) - \sum_{j=0}^{L+1}\log(1-\mathfrak{u}_{j}(C))\nonumber\\
\hspace{27mm}
-\sum_{j \in L+2-P^{+}} \Big(\log(1-\mathfrak{u}_{j}^{*}(C)) - \log(1-\mathfrak{u}_{j}(C)\Big)\\
\hspace{35mm}
-\sum_{j \in -P^{-}} \Big(\log(1-\mathfrak{u}_{j}^{*}(C)) - \log(1-\mathfrak{u}_{j}(C)\Big)\;.\nonumber
\end{eqnarray}
The KPZ regime corresponds for any $P$ to a value of the parameter $C$ close to the critical value $C_{*}\simeq-\rme\,L/4^{L+2}$ when $L\to\infty$.

The large $L$ asymptotics of the extensive sums between $0$ and $L+1$ were obtained in \cite{CN2018.1} for the study of stationary large deviations, and we recall the results in the next section. The  large $L$ asymptotics of the finite sums are then considered in the following section. Putting these results together, we recover (\ref{vp_eq1}), (\ref{vp_eq2}) with $\chi_{P}$ defined by (\ref{chi_P}).
\end{subsection}

\begin{subsection}{Large \texorpdfstring{$L$}{L} asymptotics of sums over the filled Fermi sea}
\label{bulk_asymptotics}
Given a function $h$ analytic in the vicinity of the Bethe roots $\mathfrak{u}_{j}(C)$, sums of the $h(\mathfrak{u}_{j}(C))$ can in principle be evaluated by residues using
\begin{equation}
\sum_{j=0}^{L+1}h(\mathfrak{u}_{j}(C))=\oint\frac{\rmd u}{2\rmi\pi}\,\frac{h(u)\,\partial_{u}R(u,C)}{R(u,C)}\;,
\end{equation}
where the contour encircles only the roots $\mathfrak{u}_{j}(C)$, $j=0,\ldots,L+1$ of $R(u,C)=0$. From (\ref{u C->0}), these roots are distinguished among the $2L+4$ roots $u$ of $R(u,C)=0$ as the only ones that converge to $0$ when $C\to0$. It is thus possible to compute the residues perturbatively in $C$, leading to a series of the form
\begin{equation}
\sum_{j=0}^{L+1}h(\mathfrak{u}_{j}(C))=\sum_{k=0}^{\infty}d_{k}\,C^{k}\;,
\end{equation}
with coefficients $d_{k}$ that depend on $L$ and on the function $h$. At large $L$, the coefficients $d_{k}$ may simplify so that the sum over $k$ can then be evaluated explicitly.

This method was used to compute stationary large deviations of the current for TASEP with periodic \cite{DL1998.1} and open \cite{CN2018.1} boundary conditions using Bethe ansatz. Since the stationary state precisely corresponds to the choice of the Bethe roots $\mathfrak{u}_{j}(C)$, the results at $\alpha=\beta=1$ from \cite{CN2018.1} are readily usable here. For more general $\alpha,\beta$ in the maximal current phase, we refer to \cite{LM2011.1}, where similar contour integrals were considered in a matrix product approach to stationary large deviations.

In our notations, taking
\begin{equation}
\label{C[v]}
C=\rme^{v}\,|C_{*}|\;,
\end{equation}
we obtain the large $L$ asymptotics
\begin{equation}
\fl\hspace{5mm}
\sum_{j=0}^{L+1}\frac{\mathfrak{u}_{j}(C)}{1-\mathfrak{u}_{j}(C)}\simeq
\alpha+\beta-2
+\frac{\chi'(v)}{\sqrt{L}}
-\Big(\frac{2a}{(1-a)^{2}}+\frac{2b}{(1-b)^{2}}\Big)\frac{\chi'(v)}{L^{3/2}}
+\frac{3\chi(v)}{8L^{3/2}}\;.
\end{equation}
and
\begin{equation}
\fl\hspace{1mm}
\sum_{j=0}^{L+1}\log(1-\mathfrak{u}_{j}(C))\simeq
-\log(\alpha\beta)-\frac{2\chi'(v)}{\sqrt{L}}
+\Big(2+\frac{4a}{(1-a)^{2}}+\frac{4b}{(1-b)^{2}}\Big)\frac{\chi'(v)}{L^{3/2}}
+\frac{\chi(v)}{4L^{3/2}}\;,
\end{equation}
where $\log$ is understood with branch cut $\mathbb{R}^{-}$, and the function $\chi$ is defined in (\ref{chi_integral}).

For finite $L$, the two sums above are analytic in $\mathbb{C}\setminus\mathbb{R}^{-}$ since this is the domain of analyticity of the functions $\mathfrak{u}_{j}(C)$, and $1-\mathfrak{u}_{j}(C)$ never crosses the branch cut $\mathbb{R}^{-}$ of the logarithm, see figure~\ref{fig uj}. At large $L$, with the definition (\ref{C[v]}) for the parameter $v$, the asymptotic expressions obtained must be periodic in $v$ with period $2\rmi\pi$, giving indeed the function $\chi$ with horizontal branch cuts. The fact that $\chi_{\emptyset}$ with vertical cuts appears instead in the final result (\ref{chi_P}) comes merely from a different choice of labelling of the eigenstates with $|\Im~v|>\pi$, as explained in section~\ref{section a.c.}.
\end{subsection}

\begin{subsection}{Large \texorpdfstring{$L$}{L} asymptotics for the excitations}
Taking $C$ as in (\ref{C[v]}) and writing $u=-1-\frac{\lambda}{\sqrt{L}}$, we obtain at leading order in $L$ from $R(u,C)=0$ the identity $\frac{\lambda^{2}}{4}\,\rme^{\lambda^{2}/4}=\rme^{-v-1}$, which implies that Bethe roots close to $-1$ at large $L$ may be expressed in terms of the square root Lambert functions $y_{j}$ defined in (\ref{yj}). A careful treatment gives $1+\mathfrak{u}_{j}(C)\simeq-2\rmi y_{-j}(v)/\sqrt{L}$ and $1+\mathfrak{u}_{j}^{*}(C)\simeq2\rmi y_{-j}(v)/\sqrt{L}$ for finite $j>0$, and $1+\mathfrak{u}_{L+2+j}(C)\simeq-2\rmi y_{-j}(v)/\sqrt{L}$ and $1+\mathfrak{u}_{L+2+j}^{*}(C)\simeq2\rmi y_{-j}(v)/\sqrt{L}$ for finite $j<0$.

Pushing the expansion up to order $L^{-3/2}$, we obtain for finite $j<0$ the asymptotics
\begin{eqnarray}
\fl\hspace{1mm}
\mathfrak{u}_{-j}(C)\simeq
-1-\frac{2\rmi y_{j}(v)}{\sqrt{L}}
+\frac{2y_{j}(v)^{2}}{L}
+\frac{3\rmi y_{j}(v){^3}}{2L^{3/2}}
+\frac{\rmi y_{j}(v)}{L^{3/2}}\Big(\frac{3}{2}+\frac{4a}{(1-a)^{2}}+\frac{4b}{(1-b)^{2}}\Big)\\
\fl\hspace{1mm}
\mathfrak{u}_{-j}^{*}(C)\simeq
-1+\frac{2\rmi y_{j}(v)}{\sqrt{L}}
+\frac{2y_{j}(v)^{2}}{L}
-\frac{3\rmi y_{j}(v){^3}}{2L^{3/2}}
-\frac{\rmi y_{j}(v)}{L^{3/2}}\Big(\frac{3}{2}+\frac{4a}{(1-a)^{2}}+\frac{4b}{(1-b)^{2}}\Big)\;,
\end{eqnarray}
and exactly the same expressions with finite $j>0$ respectively for $\mathfrak{u}_{L+2-j}(C)$ and $\mathfrak{u}_{L+2-j}^{*}(C)$. These expansions allow to compute the asymptotics of the terms in the sums over $L+2-P^{+}$ and $-P^{-}$ in (\ref{sum_E_asymptotic}), (\ref{sum_gamma_asymptotic}), from which $\chi_{P}(v)$ and $\chi_{P}'(v)$ are recovered after combining with the results of the previous section.
\end{subsection}

\end{section}

\begin{section}{Conclusions}
Exact expressions (\ref{vp_eq1})-(\ref{vp_eq2}) were obtained in this paper for spectral gaps of open TASEP in the maximal current phase. An important finding is that all the gaps may be written in a parametric way as $E\sim\chi_{P}(v)$ with $v$ solution of $\chi_{P}'(v)$, where the functions $\chi_{P}$ are branches of an analytic functions defined on an infinite genus Riemann surface $\mathcal{R}$. In particular, all the gaps may be obtained by analytic continuations from the known stationary eigenvalue \cite{GLMV2012.1} describing large deviations of the current in the stationary state. It would be interesting to understand whether this is a general feature for spectral gaps of Markov processes, or merely a consequence of the integrability of the model. A clear physical interpretation of the parameter $v$ would be especially welcome.

The calculation of the gaps is a first step toward an exact description of the finite volume KPZ fixed point on an interval. The Riemann surface $\mathcal{R}$ generated by the function $\chi$ (\ref{chi_integral}) is expected to play the same role as the Riemann surface generated by half-integer polylogarithms plays for the KPZ fixed point with periodic boundaries \cite{P2020.1}. The contribution of eigenvectors and their scalar product with initial states will of course be needed in order to compute the relaxation dynamics of KPZ fluctuations on the interval.

An interesting issue is whether the KPZ fixe points with periodic and open boundaries are connected with one another. It is known that at the points $\alpha=1/2,\beta=1$ and $\alpha=1,\beta=1/2$, which are on the boundary of the maximal current phase, the spectrum of open TASEP with $L$ sites is a subset of the spectrum of periodic TASEP with $2L+2$ sites and $L+1$ particles \cite{GLMV2012.1}. This means that in the vicinity of the boundary of the maximal current phase, the Riemann surface describing spectral gaps should interpolate between $\mathcal{R}$ and the corresponding Riemann surface for KPZ with periodic boundaries.
\end{section}

\appendix
\begin{section}{Richardson extrapolation}
\label{apendix num}
In order to extract asymptotic values $L\to\infty$ from finite size Bethe ansatz results, we use Richardson extrapolation with the Bulirsch-Stoer algorithm \cite{HS1988.1}. The method allows for a very precise determination of limit values with a reliable estimation of the error from relatively few finite size values, if the quantity one is interested in has an expansion of the form
\begin{equation}
A(h) = A^* - \sum_{k=1}^{\infty} a_k h^{\theta k}\;.
\end{equation}
This is in particular the case for the eigenvalues $L^{3/2}E$ of $M(\gamma)$ in the KPZ regime, with $h=1/L$ and the exponent $\theta=1$.

Starting from a finite sequence of values $h_0 < \ldots < h_n$, with $A_i^{(0)}=A(h_i)$, the Bulirsch-Stoer algorithm consists in building recursively the array
\begin{equation}
\begin{array}{ccccc}
A_0^{(0)} & A_1^{(0)} & \cdots & A_{n-1}^{(0)} & A_n^{(0)}\\[2mm]
A_0^{(1)} & A_1^{(1)} & \cdots & A_{n-1}^{(1)} &\\[2mm]
\cdots & \cdots & \cdots &&\\[2mm]
A_0^{(n)} &&&&
\end{array}
\end{equation}
with
\begin{equation}
A_m^{(p)} = A_{m+1}^{(p-1)} + \frac{A_{m+1}^{(p-1)} - A_m^{(p-1)}}{ \big( \frac{h_{m+1}}{h_{m+p+1}}\big)^\theta \bigg(1 - \frac{A_{m+1}^{(p-1)} - A_m^{(p-1)}}{A_{m+1}^{(p-1)} - A_{m+1}^{(p-2)} - 1} \bigg)}\;.
\end{equation}
The extrapolated value for $A^*$ is then $A_0^{(n)}$, and the quantity $|A_1^{(n)}-A_1^{(n-1)} |+|A_1^{(n)}-A_2^{(n-1)}|+|A_1^{(n-1)}-A_2^{(n-1)}|$ is a good estimation for the order of magnitude of the error.
\end{section}

\begin{section}{Orbits under the action of \texorpdfstring{$T_r$}{Tr}}
\label{appendix Tr}
In this appendix we prove two results used in section~\ref{section a.c.} about orbits of finite sets of integers under the action of the operator $T_r$.
 
\begin{subsection}{Existence of a set \texorpdfstring{$P^*$}{P*} with \texorpdfstring{$|P_{+}^{*}|=|P_{-}^{*}|$}{|P+|=|P-|} in each orbit}
Let $P$ be a finite set of integers, with positive elements $P_{+}=\{k^+_1,\ldots,k^+_h\}$, $0<k^+_1<\ldots<k^+_h$ and negative elements $P_{-}=\{k^-_\ell,\ldots,k^-_1\}$, $k^-_\ell<\ldots<k^-_1<0$. Then, the action on $P$ of the operator $T_r$ from (\ref{Tr P}) writes
\begin{equation}
\label{action_Tr}
\fl\hspace{5mm}
T_r P = \left\{\begin{array}{ll}
	\{k^-_\ell + 1,..., k^-_2 +1, k^+_1 +1,...,k^+_h+1\} & \text{if} \quad 0 \in P \; \& \; -1 \in P \\
	\{k^-_\ell + 1,..., k^-_1 +1, 1, k^+_1 +1,...,k^+_h+1\} & \text{if} \quad 0 \in P \; \& \; -1 \notin P \\
	\{k^-_\ell + 1,..., k^-_2 +1, 0, k^+_1 +1,...,k^+_h+1\} & \text{if} \quad 0 \notin P \; \& \; -1 \in P \\
	\{k^-_\ell + 1,..., k^-_1 +1, 0,1, k^+_1 +1,...,k^+_h+1\} & \text{if} \quad 0 \notin P \; \& \; -1 \notin P \\
\end{array}\right.\;.
\end{equation}
We observe that in any case, the application of $T_r$ increases the excess number of positive elements in $P$ by exactly $1$, so that for any $P$, the set $P^*=T_r^{|P_-|-|P_+|} P$ is the unique element with $|P_+^*|=|P_-^*|$ in the orbit of $P$ under the action of $T_r$.
\end{subsection}

\begin{subsection}{\texorpdfstring{$P^{*}$}{P*} does not contain \texorpdfstring{$0$}{0} for analytic continuations from \texorpdfstring{$P=\emptyset$}{P={}}}
As in the previous section, we denote by $P^*$ the unique representative with $|P_+^*|=|P_-^*|$ in the orbit of any finite set $P\subset\mathbb{Z}$ under the action of $T_r$. We want to show that for any $P$ such that the function $\chi_P$ may be obtained from $\chi_\emptyset$ by analytic continuations, i.e. any $P$ that can be obtained from the empty set $\emptyset$ by repeated action of the operators $A_n^{\text{l|r}}$, $n\in\mathbb{Z}$, the corresponding set $P^{*}$ does not contain $0$.

Since by construction $(T_r^{n} P)^{*}=P^{*}$, the identities
\begin{eqnarray}
&& A_{n}^{\text{l}}=T_{\text{r}}^{-n} T_{\text{l}}^{n}\\
&& A_{n}^{\text{r}}=T_{\text{l}}^{-n} T_{\text{r}}^{n}\;,\nonumber
\end{eqnarray}
imply that we only need to prove that the collection of all $P \subset \mathbb{Z}$ such that $0\notin P^*$ is globally stable under the action of $T_\text{l}$, defined by $T_{\text{l}}^{n}P=P-n$. But we observe from (\ref{action_Tr}) that $0 \in P^*$ is equivalent to $0\in P$ (respectively $0 \notin P$) if $|P_+|-|P_-|$ is even (resp. odd). The fact that under the replacement $P\to T_\text{l}P$, $|P_+|-|P_-|$ is left unchanged if $0\notin P$ and is decreased by one otherwise concludes the proof.
\end{subsection}

\end{section}

\vspace{10mm}

\end{document}